\documentclass[fleqn,usenatbib]{mnras}
\usepackage{pgfplotstable,filecontents}
\usepackage{newtxtext,newtxmath}
\usepackage{csvsimple}
\usepackage{soul,ulem}
\usepackage[T1]{fontenc}

\DeclareRobustCommand{\VAN}[3]{#2}
\let\VANthebibliography\thebibliography
\def\thebibliography{\DeclareRobustCommand{\VAN}[3]{##3}\VANthebibliography}

\usepackage{graphicx}	% Including figure files
\usepackage{amsmath}	% Advanced maths commands
\newcommand{\kms}{km~s$^{-1}$}
\newcommand{\Msun}{M$_{\odot}$}
\newcommand{\Rc}{$R_{\mathrm{c}}$}
\newcommand{\rms}{$\sigma_{\mathrm{rms}}$}
\newcommand{\siggs}{$\sigma_{\mathrm{gs,los}}$}
\newcommand{\sigobs}{$\sigma_{\mathrm{obs,los}}$}
\newcommand{\Mgas}{$M_{\mathrm{gas}}$}

\newcommand{\Xco}{$X_{\mathrm{CO}}$}
\newcommand{\surfgas}{$\Sigma_{\mathrm{gas}}$}

\newcommand{\virialpha}{$\alpha_{\mathrm{obs,vir}}$}
\newcommand{\Lco}{$L_{\mathrm{CO(1-0)}}$}
\newcommand{\Fco}{$F_{\mathrm{CO(1-0)}}$}
\newcommand{\CO}{$^{12}$CO(1-0)}

\newcommand{\punit}{K~cm$^{-3}$}

\title[WISDOM XXI.\ GMCs in NGC~613]{WISDOM Project -- XXI.\ Giant molecular clouds in the central region of the barred spiral galaxy NGC~613: a steep size -- linewidth relation}

\author[Choi et al.]{Woorak Choi,$^{1}$\thanks{E-mail: woorak.c@gmail.com}
  Martin Bureau,$^{2}$\thanks{E-mail: martin.bureau@physics.ox.ac.uk}
  Lijie Liu,$^{3,4}$
  Michele Cappellari,$^{2}$
  Timothy A.\ Davis,$^{5}$
  \newauthor
  Jindra Gensior,$^{6}$
  Fu-Heng Liang,$^{2}$
  Anan Lu,$^{7}$
  Sanghyuk Moon,$^{8}$
  Ilaria Ruffa,$^{5,9}$
  \newauthor
   Thomas G.\ Williams$^{2}$ and Aeree Chung$^{1}$\thanks{E-mail: achung@yonsei.ac.kr}
  \\
  $^{1}$Department of Astronomy, Yonsei University, 50 Yonsei-ro, Seodaemun-gu, Seoul 03722, Republic of Korea\\
  $^{2}$Sub-department of Astrophysics, Department of Physics, University of Oxford, Keble Road, Oxford OX1~3RH, UK \\
  $^{3}$Cosmic Dawn Center (DAWN), Technical University of Denmark, DK2800 Kgs.\ Lyngby, Denmark \\
  $^{4}$DTU-Space, Technical University of Denmark, Elektrovej 327, DK2800 Kgs.\ Lyngby, Denmark\\
  $^{5}$Cardiff Hub for Astrophysics Research and Technology, School of Physics and Astronomy, Cardiff University, Queens Buildings, Cardiff, CF24~3AA, UK\\
  $^{6}$Department of Astrophysics, University of Zurich, Winterthurerstrasse 190, 8057 Z\"{u}rich, Switzerland\\
  $^{7}$McGill Space Institute and Department of Physics, McGill University, 3600 University Street, Montreal QC, H3A~2T8, Canada\\
  $^{8}$Department of Astrophysical Sciences, Princeton University, Princeton, NJ 08544, USA\\
  $^{9}$INAF - Istituto di Radioastronomia, via P.Gobetti 101, 40129 Bologna, Italy
}

\date{Accepted XXX. Received YYY; in original form ZZZ}

\pubyear{2023}

\begin{document}
\label{firstpage}
\pagerange{\pageref{firstpage}--\pageref{lastpage}}
\maketitle

% Abstract of the paper
\begin{abstract}
    NGC~613 is a nearby barred spiral galaxy with a nuclear ring. Exploiting high spatial resolution ($\approx20$~pc) Atacama Large Millimeter/sub-millimeter Array \CO\ observations, we study the giant molecular clouds (GMCs) in the nuclear ring and its vicinity, identifying $158$ spatially- and spectrally-resolved GMCs. The GMC sizes ($R_{\mathrm{c}}$) are comparable to those of the clouds in the Milky Way (MW) disc, but their gas masses, observed linewidths ($\sigma_{\mathrm{obs,los}}$) and gas mass surface densities are larger. The GMC size -- linewidth relation ($\sigma_{\mathrm{obs,los}}\propto R_{\mathrm{c}}^{0.77}$) is steeper than that of the clouds of the MW disc and centre, and the GMCs are on average only marginally gravitationally bound (with a mean virial parameter $\langle\alpha_{\mathrm{obs,vir}}\rangle\approx1.7$). We discuss the possible origins of the steep size -- linewidth relation and enhanced observed linewidths of the clouds and suggest that a combination of mechanisms such as stellar feedback, gas accretion and cloud-cloud collisions, as well as the gas inflows driven by the large-scale bar, may play a role.  
\end{abstract}

%%%%%%%%%%%%%%%%%%%%%%%%%%%%%%%%%%%%%%%%%%%%%%%%%%

%%%%%%%%%%%%%%%%% BODY OF PAPER %%%%%%%%%%%%%%%%%%

\section{Introduction}
\label{sec:intro}

As giant molecular clouds (GMCs) are the gas reservoirs where most star formation takes place, understanding their life cycle and characteristics is essential to comprehend the formation and evolution of galaxies. Early studies of GMCs were limited to Milky Way (MW) and Local Group galaxies such as the Large Magellanic Cloud (LMC; e.g.\ \citealt{fukui2008}), Small Magellanic Cloud (SMC; e.g.\ \citealt{muller2010}), M~31 \citep[e.g.][]{rosolowsky2007} and M~33 \citep[e.g.][]{rosolowsky2003, rosolowsky2007etal}. These studies revealed that the GMCs of these galaxies have similar properties and size -- linewidth relations, with typical sizes of $10$ -- $100$~pc, typical masses of $10^4$ -- $10^7$~\Msun\ and typical velocity dispersions of $1$ -- $10$~\kms \citep[e.g.][]{larson1981, bolatto2008}.

With improvements of the resolution and sensitivity of molecular line observations, GMCs are now studied in extragalactic objects, revealing deviations from the properties of Local Group GMCs \citep[e.g.][]{bolatto2008, rosolowsky2021}. For example, the characteristics of the GMCs of some late-type galaxies (LTGs) vary with galactic environment and do not universally conform to the usual scaling relations of MW clouds (e.g.\ M~51, \citealt{hughes2013}, \citealt{colombo2014m51}; NGC~253, \citealt{Leroy2015cpropstoo}). The GMCs of early-type galaxies (ETGs) also have either no or different size -- linewidth relations and are brighter, denser and have higher velocity dispersions than GMCs in the MW disc (MWd) and Local Group galaxies \citep{utomo2015, liu2021ngc4429}. In addition, the molecular gas (and clouds) in the bar regions of local star-forming galaxies has higher velocity dispersions and surface densities than that in the discs \citep{sun2020_bar_phangs, sun2022_phangs_bar}. These results strongly suggest that galactic environment affects GMC properties. More studies of GMCs in galaxies with diverse morphologies and sub-structures are thus required to probe these variations and unravel the underlying physics.

Barred galaxies generally have gas inflows toward their central regions, due to their non-axisymmetric gravitational potentials \citep[e.g.][]{anthanassoula1992_dust_shock, sormani2015_bar_gasinflow,sormani2019_gasinflow_obs,henshaw2023_cmz_review,sormani2023_ngc1097_gasinflow}. Several CO surveys have shown higher concentrations of molecular gas in the central regions of barred galaxies than in unbarred galaxies \citep[e.g.][]{sakamoto1999_bar_concentration}, as well as increased molecular gas linewidths and surface densities \citep{sun2020_bar_phangs, sun2022_phangs_bar}. Recent high-resolution CO observations of barred disc galaxies have also revealed the presence of distinct structures similar to those observed at optical wavelengths, such as nuclear rings, nuclear bars and nuclear spiral arms, with non-circular motions \citep[e.g.][]{salak2016,cuomo2019_bar_pattern,BewketuBelete2021,sato2021}. Studying barred disc galaxies therefore enables the exploration of GMC properties, including scaling relations, within different galactic environments. However, there have been only a few studies investigating GMCs in barred galaxies so far \citep[e.g.][]{hirota2018_m83_bar, maeda2020a_n1300_bar, sato2021}, including GMCs in the MW \citep[e.g.][]{solomon1987,oka1998,heyer2009,kauffmann2017}. \citet{rosolowsky2021} investigated GMCs in ten nearby galaxies, but only one out of their six barred galaxies was observed with a spatial resolution sufficient to spatially resolve individual GMCs ($<30$~pc spatial resolution) and thus analyse their scaling relations and physical properties such as size, mass and linewidth.

\citet{choi2023} recently investigated the molecular gas of the barred spiral galaxy NGC~5806 at the GMC scale ($\approx24$~pc spatial resolution). Despite the GMCs having typical properties on average, they reported that the GMCs in the nuclear ring and its vicinity have a size ($R_{\mathrm{c}}$) -- linewidth ($\sigma_{\mathrm{obs,los}}$) relation ($\sigma_{\mathrm{obs,los}}\propto R_{\mathrm{c}}^{1.2}$) that is much steeper than that of the GMCs in the MW or other galaxies. This result implies that the GMCs at the centres of barred galaxies can differ from those in the discs.
However, as this behaviour has been observed in a single object and there is no adequate sample for comparison, one cannot conclude yet that it is the bar and associated gas inflow toward the nuclear ring that cause the steep size -- linewidth relation.

To further test this conjecture, we examine the GMCs at the centre of the barred spiral galaxy NGC~613 (SBbc(rs)) in this paper. NGC~613 has a large-scale bar, an inner star-forming ring encircling the bar and weak spiral arms protruding from the bar, at a distance $D=17.5$~Mpc \citep{tully1988_distance}. In the central regions (i.e.\ well within a diameter of $1$~kpc), NGC~613 has a bright core and a well-studied star-forming nuclear ring \citep{hummel1992, boker2007, boker2008, falcon-barroso2008, falcon-barroso2014, miyamoto2017_ngc613, miyamoto2018_ngc613}, that are prominent in both optical continuum and molecular gas emission. NGC~613 also harbours a nuclear spiral \citep{audibert2019}, a low-luminosity active galactic nucleus (LL-AGN; \citealt{veron-cetty_1986,hummel1987,goulding2009}), prominent radio jets \citep{miyamoto2017_ngc613}, shock excitation \citep{falcon-barroso2014} and a shock co-spatial with the peak of the radio jet \citep{davies2019_ngc613_shock}.
The size of the nuclear ring of NGC~613 (radius $300$~pc) is almost identical to that of NGC~5806 (radius $330$~pc), and the molecular gas mass of NGC~613 is about $30$\% larger than that of NGC~5806 when a Galactic conversion factor is applied. NGC~613 also has a total stellar mass ($\approx 4.5\times10^{10}$~\Msun) similar to that of NGC~5806 ($\approx 3.9\times10^{10}$~\Msun), and a relative size of the nuclear ring with respect to the galaxy optical diameter $D_{25}$ ($0.024$) similar to that of NGC~5806 ($0.023$; \citealt{aguero2016_nuclear_ring}). This leads us to expect the nuclear ring of NGC~613 to have conditions similar to those of the NGC~5806 nuclear ring, making a comparison highly relevant. \citet{sato2021} recently investigated the properties of the molecular gas in and around the nuclear ring of NGC~613. While they identified GMCs and measured their properties, they did not conduct a detailed analysis of the GMC characteristics, such as the size -- linewidth relation.
We thus first identify and then thoroughly explore the properties of the GMCs in and around the nuclear ring of NGC~613 here. Key characteristics of NGC~613 are summarised in \autoref{tab:ngc613_key}.

\begin{table}
  \caption{NGC~613 characteristics.}
  \label{tab:ngc613_key}
  \resizebox{\columnwidth}{!}{
    \begin{tabular}{lcc}
      \hline
      Quantity & Value & Reference \\
      \hline
      Type & SB(rs)bc &  (1) \\
      R.A.\ (J2000) & $1^{\mathrm{h}}34^{\mathrm{m}}18\fs19$ & (2) \\
      Dec.\ (J2000) & $-29\degr25\arcmin6\farcs6$ & (2) \\
      Distance (Mpc) & $17.5$ & (3) \\
      Position angle (degree) &  $118\pm4$ & (2) \\
      Inclination (degree) & $46\pm1$ & (2) \\
      Systemic velocity (\kms) & $1471\pm3$ & (2) \\
      Central molecular gas mass (M$_{\odot}$) & $3\times10^8$ & (2) \\
      Total stellar mass (M$_{\odot}$) & $4.5\times10^{10}$ & (4) \\
      Stellar velocity dispersion (\kms) & $92\pm3$ & (5) \\
      Total star-formation rate (M$_{\odot}$~yr$^{-1}$) & $5.2$ & (4) \\
      \hline
    \end{tabular}
  }
  \begin{minipage}{\columnwidth}
    \vspace{0.1cm}
    {\it References:} (1) \citet{devaucouleurs1991}. (2) \citet{miyamoto2017_ngc613}. The central molecular gas mass was calculated using the $^{12}$CO(3-2) intensity integrated over the central region of NGC~613 only (radius $\leq22\arcsec$), assuming a CO-to-molecules conversion factor $X_{\mathrm{CO(3-2)}}=0.5\times10^{20}$~cm$^{-2}$~(K~km~s$^{-1}$)$^{-1}$. (3) \citet{tully1988_distance}. (4) \citet{combes2019_ngc613_mass}. The total stellar mass was calculated using the total $3.6$~$\micron$ luminosity and a mass-to-light ratio $M/L_{3.6\micron}=0.5$~$\mathrm{M}_\odot/\mathrm{L}_{\odot,3.6\micron}$. The total star-formation rate was calculated using the total infrared luminosity (NASA/IPAC Extragalactic Database). (5) \citet{dasilva2020b_ngc613}. Luminosity-weighted stellar velocity dispersion within the central region of NGC~613 (radius $\leq7\arcsec$).\\
  \end{minipage}
\end{table}

This paper is structured as follows. In \autoref{sec:data_identification}, we describe the data and the methodology used to identify GMCs in NGC~613. The GMCs identified and their properties are discussed in \autoref{sec:properties}. In \autoref{sec:dynamics}, we investigate and discuss the GMC size -- linewidth relation and dynamical states, and the clouds in the nuclear ring only. We summarise our findings in \autoref{sec:conclusions}.

\section{Observations}
\label{sec:data_identification}

\subsection{Data}
\label{sec:data}

NGC~613 was observed in the \CO\ line (rest frequency $115.271$~GHz) using the Atacama Large Millimeter/sub-millimeter Array (ALMA). The observations were carried out using two different $12$-m array configurations in November and December 2017 (programme 2017.1.01671.S, configurations C43-6 and C43-8, PI Miyamoto). The C43-6 configuration observations had $6500$~s on-source using $47$ antennae and baselines of $15$ -- $2500$~m, leading to a maximum recoverable scale (MRS) of $5\farcs5$. The C43-8 configuration observations had $9500$~s on-source using $46$ antennae and baselines of $110$ -- $11,000$~m, leading to a MRS of $1\farcs7$. Both configurations have a primary beam of $54\farcs4$ full-width at half-maximum (FWHM). The correlator was set up with one spectral window of $468.750$~MHz ($\approx1200$~\kms) bandwidth and $1920$ channels each of $244$~kHz ($\approx0.63$~\kms) centred on the redshifted $^{12}$CO(1-0) line. The three remaining spectral windows each have a $1.875$~GHz bandwidth with $1920$ channels of $976$~kHz and were used to observe the continuum.

\subsubsection{Data reduction}

The raw data of each configuration were calibrated and then combined using the standard ALMA pipeline version 5.6.1-8, using \textsc{Common Astronomy Software Applications} (\textsc{casa}; \citealt{mcmullin2007_casa}) version 5.6.1. Continuum emission was measured using line-free channels and subtracted from the data using the \textsc{casa} task \texttt{UVCONTSUB}. We imaged the data using Briggs weighting with a robust parameter of $0.5$ and cleaned using the \texttt{tclean} task to a depth equal to twice the root-mean-square (RMS) noise of the dirty data cube (respectively the continuum image), yielding synthesised beam FWHM of $0\farcs27\times0\farcs20$ ($\approx24\times18$~pc$^2$) with a position angle of $86\degr$ for the \CO\ line and $0\farcs29\times0\farcs21$ ($\approx26\times19$~pc$^2$) with a position angle of $86\degr$ for the continuum. We chose a pixel size of $0\farcs05$ to balance image size and spatial sampling, yielding approximately $4\times6$ pixels across the synthesised beam for the line data. We thus created a fully calibrated and cleaned cube encompassing most of the primary beam spatially, with $2$~\kms\ (binned) channels spectrally. The RMS noise of the \CO\ cube is $\sigma_{\mathrm{rms}}=0.59$~mJy~beam$^{-1}$ ($1.10$~K) per channel. The continuum image has the RMS noise of $6.5$~$\mu$Jy~beam$^{-1}$ and the flux of the continuum emission within the nuclear ring (radius $\leq6\arcsec$) is $11.7$~mJy.

\subsection{Moment maps}\label{sec:moments}
 
\begin{figure*}
  \centering
  \includegraphics[height=0.92\columnwidth]{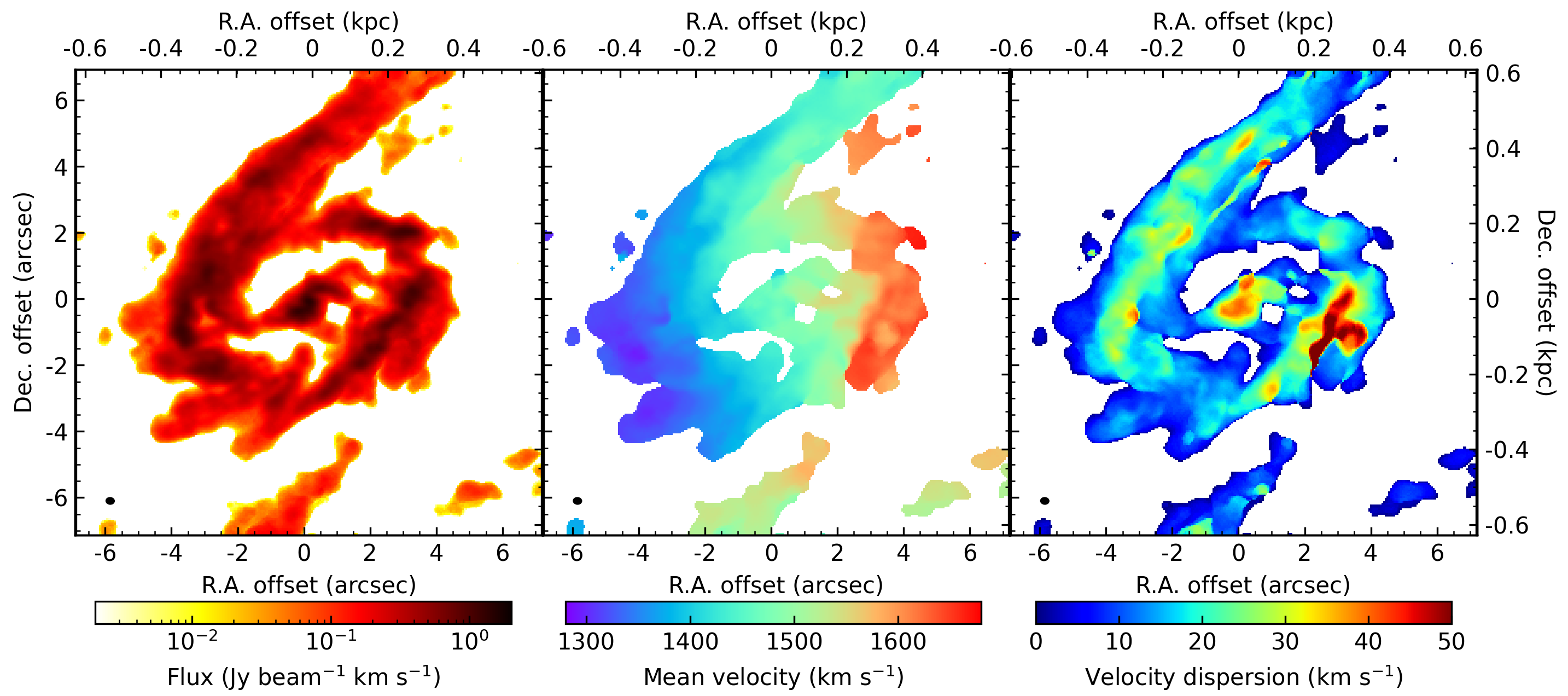}\hfill
  \includegraphics[height=0.7\columnwidth]{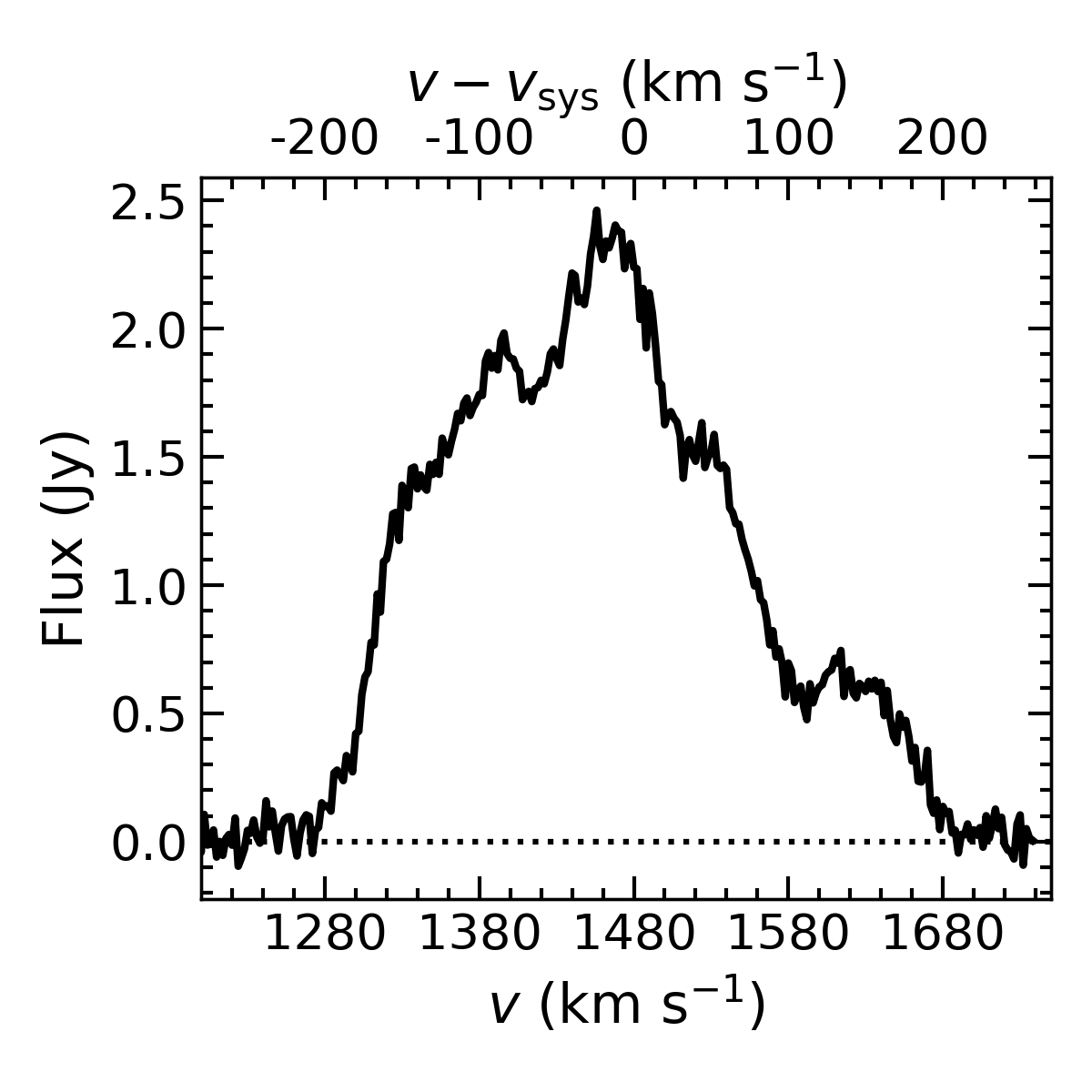}\hfill
  \includegraphics[height=0.7\columnwidth]{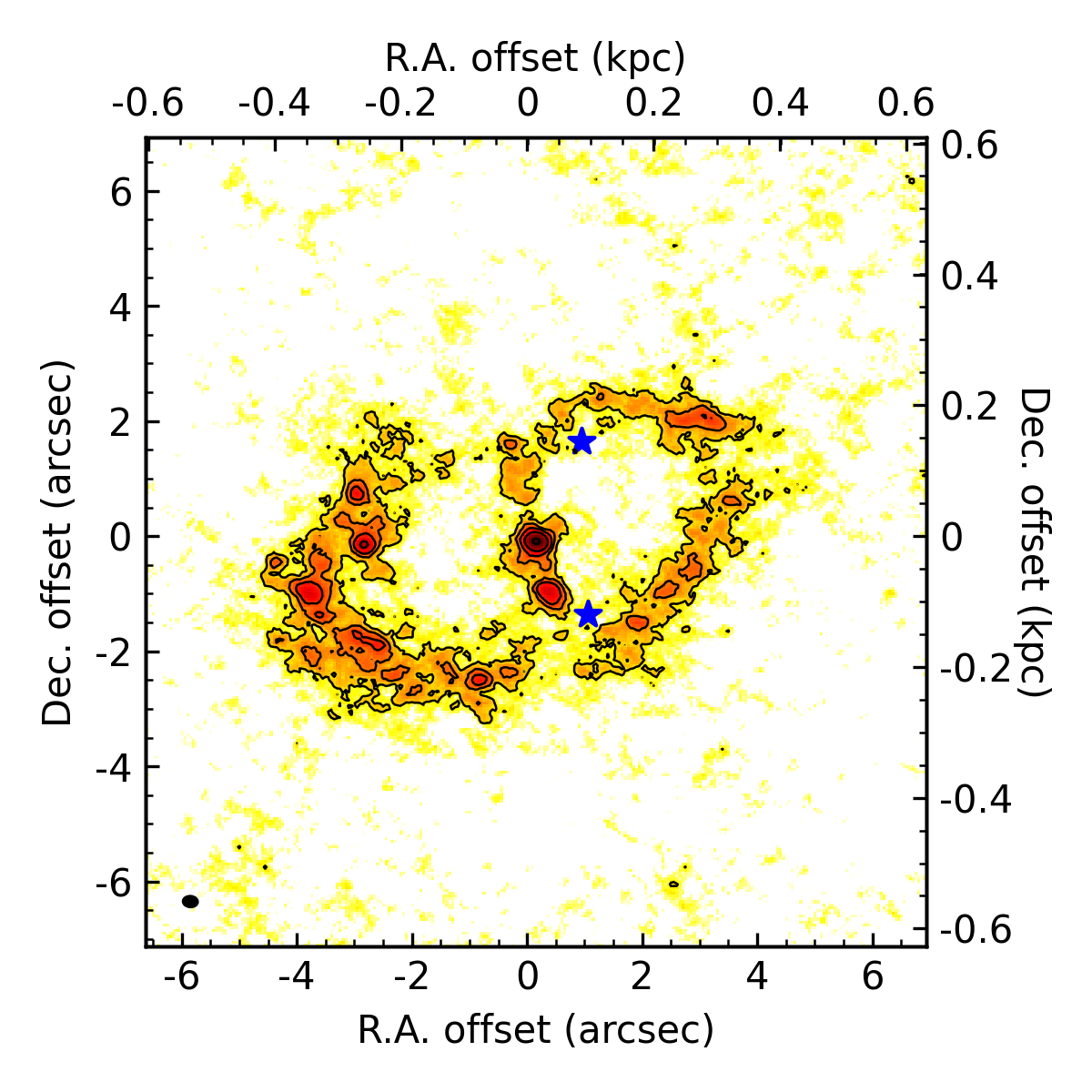}\hfill
  \includegraphics[height=0.7\columnwidth]{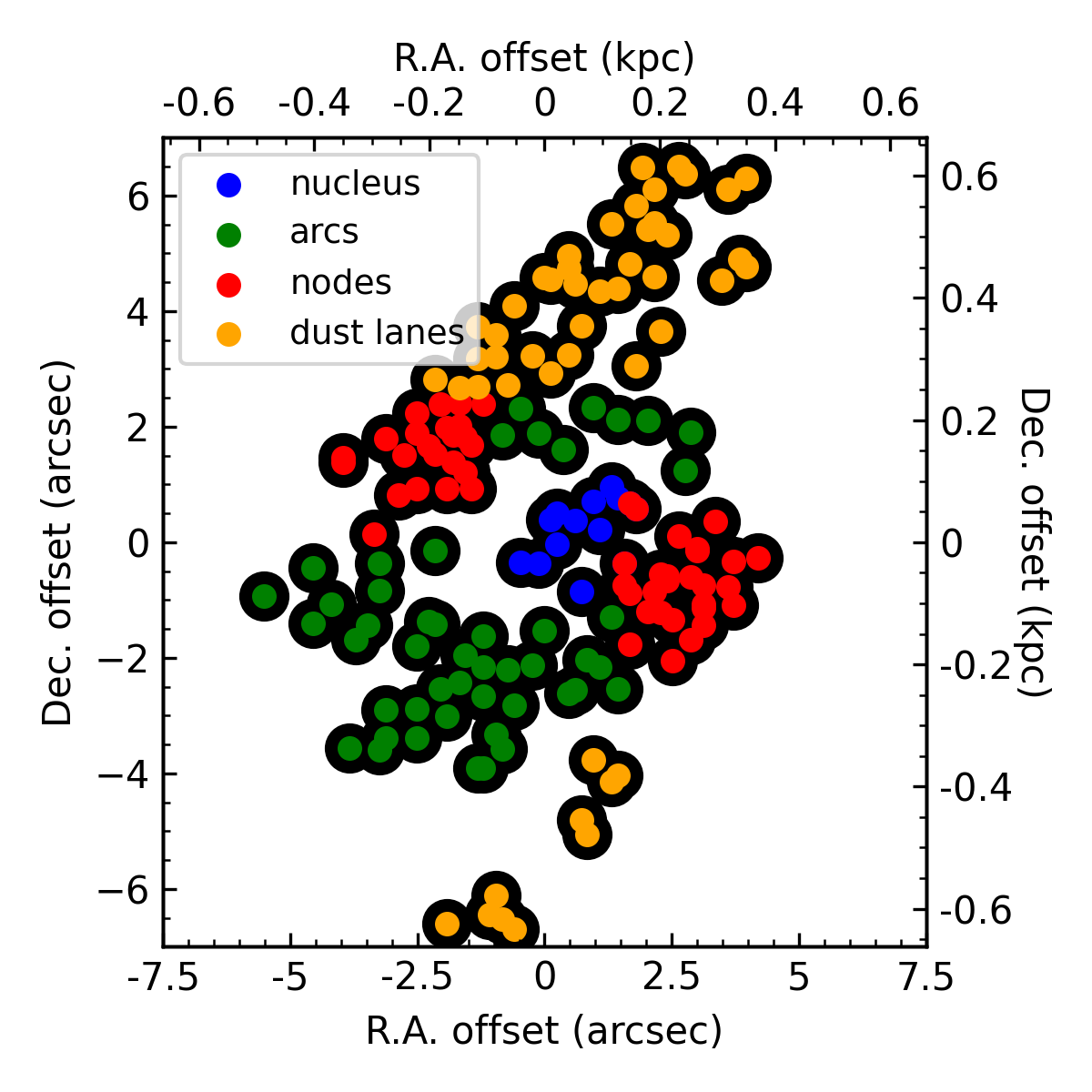}\hfill
  \caption{\CO\ emission of NGC~613. \textbf{Top-left}: zeroth-moment (integrated-intensity) map. \textbf{Top-middle}: first-moment (intensity-weighted mean line-of-sight velocity) map. \textbf{Top-right}: second-moment (intensity-weighted line-of-sight velocity dispersion) map. The synthesised beam of $0\farcs27\times0\farcs20$ ($\approx24\times18$~pc$^2$) is shown in the bottom-left corner of each moment map as a filled black ellipse. \textbf{Bottom-left}: Integrated \CO\ spectrum, extracted from a $15\arcsec\times15\arcsec$ region around the galaxy centre. \textbf{Bottom-middle}: $100$~GHz continuum map. The synthesised beam of $0\farcs29\times0\farcs21$ is shown in the bottom-left corner as a filled black ellipse. The blue stars indicate the regions where SiO(2-1) is detected \citep{miyamoto2017_ngc613}. \textbf{Bottom-right}: resolved clouds (black circles; see \autoref{fig:cloudfinding_result}) of the nucleus (blue), arcs (green), nodes (red) and dust lanes (yellow), as defined in \autoref{sec:moments}.}
  %m Note also that you could put the bottom panels in an entirely different figure, which could then use only half the width of the page. Implement?
  \label{fig:moments}
\end{figure*}

\autoref{fig:moments} shows the zeroth-moment (integrated-intensity), first-moment (intensity-weighted mean line-of-sight velocity) and second-moment (intensity-weighted line-of-sight velocity dispersion) maps of the \CO\ emission of NGC~613. These were generated using a smooth-moment masking method \citep[e.g.][]{dame2011}. In summary, we convolved the \CO\ data cube spatially with a Gaussian of width equal to that of the synthesised beam and Hanning-smoothed the cube spectrally. To create a mask we then selected only pixels with intensities exceeding three times the RMS noise of the smoothed cube, and applied this mask to the original cube to generate the moment maps.

The integrated intensity map of NGC~613 (top-left panel of \autoref{fig:moments}) reveals a highly structured molecular gas distribution. There are at least three clearly distinguishable structures: a nucleus at the very centre of the galaxy, a nuclear ring with a radius of $\approx300$~pc and offset dust lanes stretching from the northeast (toward the northwest) and the southwest (toward the southeast). The interface between the nuclear ring and the dust lane in the east has intensities higher than those in the rest of the nuclear ring, and there is a gradual decrease of the intensities as a function of the azimuthal angle in a counter-clockwise direction. The interface in the west also has higher intensities but it does not show a clear decrease of the intensities along the nuclear ring. Finally, there are faint nuclear spiral arms between the nucleus and the nuclear ring.

The mean velocity map (top-middle panel of \autoref{fig:moments}) clearly shows roughly circular motions throughout the disc and particularly within the nuclear ring, with a projected angular momentum vector pointing toward the northeast. The zero-velocity curve shows significant twists, however, particularly along the dust lanes, implying significant non-circular motions. These motions are as expected from the shocks and non-axisymmetric perturbations driven by the large-scale bar of a disc galaxy \citep[e.g.][]{anthanassoula1992_dust_shock}.

The velocity dispersion of the molecular gas (top-right panel of \autoref{fig:moments}) ranges from $0$ to $60$~\kms, higher than typical velocity dispersions across discs of galaxies but similar to those in the central region of nearby disc galaxies \citep[e.g.][]{wilson2011_vel_disp, mogotsi2016_vel_disp, sun2018_vel_disp}. The nucleus has particularly high velocity dispersion ($20$ -- $40$~\kms). As NGC~613 hosts an AGN and radio jets, this might be due to the AGN and its feedback. The velocity dispersions at the interfaces between the nuclear ring and the offset dust lanes also have high velocity dispersions ($20$ -- $60$~\kms), higher than those in the rest of the nuclear ring ($0$ -- $25$~\kms). In particular, the interface to the west of the nucleus has velocity dispersions higher ($30$ -- $60$~\kms) than those in the interface to the east of the nucleus ($10$ -- $35$~\kms), indicating that the two regions are slightly different. Indeed, \citet{sato2021} revealed that the gas inflow velocity toward the eastern side of the nuclear ring from the northeast dust lane ($\approx70$~\kms) is lower than that toward the western side of the nuclear ring from the southwest dust lane ($\approx170$~\kms). 

The \CO\ spectrum of NGC~613 integrated within a central region of $15\arcsec\times15\arcsec$ (bottom-left panel of \autoref{fig:moments}) reveals multiple peaks, suggesting again complex molecular gas distribution and kinematics. The total \CO\ flux in that region is $512$~Jy~\kms.

The $100$~GHz continuum map of NGC~613 (bottom-middle of \autoref{fig:moments}) shows two distinct structures: one associated with the nuclear ring, the other in the centre and elongated roughly north-south. By showing that the spectral index between $4.9$ and $95$~GHz $\alpha\approx-0.2$ in the nuclear ring while $\alpha\approx-0.6$ in the central region, \citet{miyamoto2017_ngc613} suggested that the continuum emission in the nuclear ring likely arise from free-free emission from \ion{H}{ii} regions while the elongated continuum emission at the centre is due to the jet. Additionally, they reported that SiO(2-1) (a shock tracer; e.g.\ \citealt{gracia-burillo_sio_shock}) is detected at both ends of the elongated continuum (blue stars in bottom-middle of \autoref{fig:moments}).

The bottom-right panel of \autoref{fig:moments} illustrates our definitions of four distinct regions based on the moment maps: nucleus (blue), arcs (green), nodes (red) and dust lanes (yellow). The nucleus encompasses the inner $180$~pc in radius. The arcs refer to the regions of relatively low velocity dispersions ($0$ -- $25$~\kms) of the nuclear ring, while the nodes refer to the regions of relatively high velocity dispersions ($\gtrsim25$~\kms) of the nuclear ring, at the interfaces between the offset dust lanes and the nuclear ring. As expected, the dust lanes encompass the two offset dust lanes stretching from the east of the nuclear ring toward the northwest and from the west of the nuclear ring toward the southeast, that are often present in barred disc galaxies \citep[e.g.][]{anthanassoula1992_dust_shock}. Note that we will also refer to the nuclear ring in its entirety, encompassing both the arcs and the nodes.

Interestingly, the molecular gas distribution of NGC~613 is very similar to that in the central region of NGC~5806 \citep{choi2023}, except for the fact that NGC~5806 has similar gas inflow velocities at the nodes on either side of the nuclear ring. The GMCs of NGC~613 and NGC~5806 are thus also expected to have similar characteristics.

\section{Cloud identification and properties}
\label{sec:properties}

\subsection{Cloud identification}
\label{sec:identification}

To identify the clouds of NGC~613 (and to be consistent with previous works), we use our own modified version of the algorithms of \textsc{cpropstoo} \citep{liu2021ngc4429}, an updated version of \textsc{CPROPS} \citep{Rosolowsky2006cprops, Leroy2015cpropstoo}. Due to fewer free parameters, our version is more efficient and leads to a more robust cloud identification in complex and crowded environments.

We refer the reader to \citet{liu2021ngc4429} for full details of our version of \textsc{cpropstoo}, but introduce here the main parameters of the algorithms. \textsc{cpropstoo} identifies and divides clouds into resolved and unresolved clouds based on fixed input parameters. The algorithm generates a three-dimensional (3D) mask using two parameters, \textit{high-threshold} and \textit{low-threshold}. The mask initially chooses only pixels for which two adjacent channels are above \textit{high-threshold}, and then expands this region to include all neighbouring pixels for which two adjacent channels are above \textit{low-threshold}. To identify individual clouds from the selected pixels, the algorithm imposes a minimum cloud area (\textit{minarea}) and a minimum number of cloud velocity channels (\textit{minvchan}). Lastly, the \textit{convexity} parameter (defined as the ratio of the volume of a cloud's 3D intensity distribution to that of its convex envelope) is used to identify structures over multiple spatial scales with less arbitrariness. We set the minimum contrast between a cloud's peak and its boundary at $\Delta T_{\mathrm{max}}=2$~$\sigma_{\mathrm{rms}}=2.20$~K.

\begin{figure}
  \centering
  \includegraphics[width=1\columnwidth]{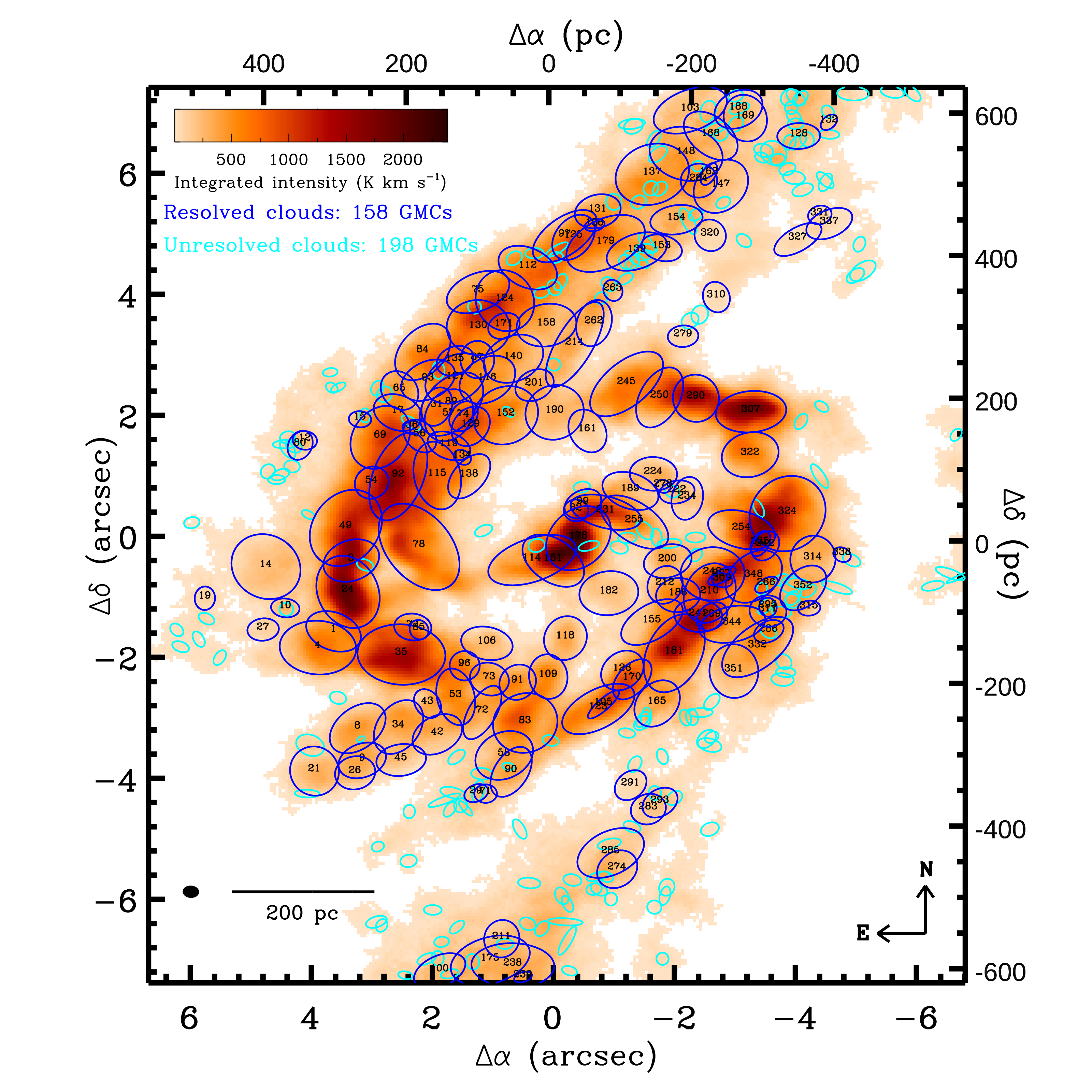}
  \caption{\CO\ integrated-intensity map of NGC~613, with the 356 GMCs identified overlaid. Resolved clouds are shown as dark blue open ellipses, unresolved clouds as cyan open ellipses. The major and minor axes of each ellipse represent the RMS spatial extent of the associated cloud along its major and minor axes, respectively, extrapolated to the limit of perfect sensitivity but not been corrected (i.e.\ deconvolved) for the finite angular resolution.} The synthesised beam ($0\farcs27\times0\farcs20$ or $\approx24\times18$~pc$^2$) is shown in the bottom-left corner as a filled black ellipse.
  \label{fig:cloudfinding_result}
\end{figure}

In this work, we run \textsc{cpropstoo} on a $15\arcsec\times15\arcsec$ region of the \CO\ cube centred on the galaxy centre, thus encompassing all of the emission detected in the zeroth-moment map (see \autoref{fig:moments}), adopting \textit{high-threshold} $=4.5$~\rms, \textit{low-threshold} $=2.0$~\rms, \textit{minvchan} $=2$ and \textit{convexity} $=0.65$. We assign \textit{minarea} a range of $120$ -- $24$ spaxels (the synthesised beam area) with a step size of $24$ spaxels in descending order, rather than adopting a single value.
% We note that $convexity$ typically ranges from $0.5$ to $0.7$.
As a result, we identify $356$ GMCs, $158$ of which are both spatially and spectrally resolved, shown in \autoref{fig:cloudfinding_result}. We note that while different sets of parameters yield slightly different cloud identification results, we have verified that for reasonable parameters the overall results and conclusions remain unaffected. 

\subsection{Cloud properties}

We calculate the physical properties of the clouds as described in \citet{liu2021ngc4429} and references therein \citep{Rosolowsky2006cprops, Leroy2015cpropstoo}. We describe here the three key properties of the clouds (size, velocity dispersion and gas mass), from which the virial parameter is derived.

The cloud size (radius \Rc) is defined as
\begin{equation}
   R_{\mathrm{c}}\equiv\eta\sqrt{\sigma_{\mathrm{maj,dc}}\,\sigma_{\mathrm{min,dc}}}\,\,\,,
\end{equation}
where $\eta$ is a geometric parameter, $\sigma_{\mathrm{maj,dc}}$ and $\sigma_{\mathrm{min,dc}}$ are the deconvolved RMS spatial extent along the major and the minor axis of the cloud, respectively, and we adopt $\eta=1.91$ for consistency with earlier studies \citep[e.g.][]{solomon1987, utomo2015, liu2021ngc4429}.

The observed velocity dispersion (\sigobs) is calculated as
\begin{equation}
   \sigma_{\mathrm{obs,los}}\equiv\sqrt{\left(\sigma_{\mathrm{v}}^2-(\Delta V_{\mathrm{chan}}^2/2\pi)\right)}\,\,\,,
\end{equation}
where $\sigma_{\mathrm{v}}$ is the second moment extrapolated to the limit of infinite sensitivity ($T_{\rm edge}=0$~K) along the velocity axis and $\Delta V_{\mathrm{chan}}$ is the channel width of $2$~\kms.

The molecular gas mass (\Mgas) is obtained from the \CO\ luminosity 
\begin{equation}
  \left(\frac{L_{\mathrm{CO(1-0)}}}{\mathrm{K~km~s^{-1}~pc^2}}\right)=\left(\frac{3.25\times10^7}{(1+z)^3}\right)\left(\frac{F_{\mathrm{CO(1-0)}}}{\mathrm{Jy~km~s^{-1}}}\right)\left(\frac{\nu_{\mathrm{obs}}}{\mathrm{GHz}}\right)^{-2}\left(\frac{D}{\mathrm{Mpc}}\right)^2\,\,\,,
\end{equation}
where \Fco\ is the \CO\ flux, $z$ the galaxy redshift and $\nu_{\mathrm{obs}}$ the observed line frequency. To obtain the molecular gas mass, we adopt a CO-to-molecule conversion factor (including helium) $X_{\mathrm{CO}}=2\times10^{20}$~cm$^{-2}$~(K~\kms)$^{-1}$, equivalent to a conversion factor $\alpha_{\mathrm{CO(1-0)}}\approx4.4$~M$_{\odot}$~(K~km~s$^{-1}$~pc$^2$)$^{-1}$, yielding
\begin{equation}
  \begin{split}
    \left(\frac{M_{\mathrm{gas}}}{\mathrm{M_\odot}}\right) & \approx\ 4.4\,\left(\frac{L_{\mathrm{CO(1-0)}}}{\mathrm{K~km~s^{-1}~pc^2}}\right)\,\,\,.\\
  \end{split} 
\end{equation}

\subsubsection{General properties}

\autoref{tab:table1} lists the three aforementioned (intensity-weighted) properties of each cloud and many more  \citep[see][]{liu2021ngc4429}: each cloud's central position (R.A.\ and Dec.), mean local standard of rest velocity ($V_{\mathrm{LSR}}$), size (radius \Rc), observed velocity dispersion (\sigobs), \CO\ luminosity (\Lco), molecular gas mass (\Mgas), peak intensity ($T_{\mathrm{max}}$), gradient-subtracted velocity dispersion (\siggs) and deprojected distance from the galaxy centre ($R_{\mathrm{gal}}$). \siggs\ represents the "turbulent" motions within a cloud, having subtracted all bulk motions due to e.g.\ shear or internal rotation, and we refer readers to \citet{utomo2015} for its definition (see also \citealt{liu2021ngc4429}). The uncertainties of the above quantities are derived via $500$ bootstrap resampling. The uncertainty of the distance to NGC~613 is not propagated through, as a distance error translates to a systematic scaling of some of the estimated quantities: $R_{\mathrm{c}}\propto D$, $L_{\mathrm{CO(2-1)}}\propto D^2$, $M_{\mathrm{gas}}\propto D^2$, $M_{\mathrm{vir}}\propto D$ and $\alpha_{\mathrm{vir}}\propto D^{-1}$ (see \autoref{sec:dynamics}).

\begin{table*}
  \centering
  \caption{Measured properties of the clouds of NGC~613. A complete machine-readable version of this table is available in the online journal version.}
  \label{tab:table1}
  \resizebox{\textwidth}{!}{
    \begin{tabular}{ccccccccccc}
      \hline
      ID & R.A. (2000) & Dec.\ (2000) & $V_{\mathrm{LSR}}$  & $R_{\mathrm{c}}$ & $\sigma_{\mathrm{obs,los}}$ & $\sigma_{\mathrm{gs,los}}$ & $L_{\mathrm{CO(2-1)}}$ & $M_{\mathrm{gas}}$ & $T_{\mathrm{max}}$ & $R_{\mathrm{gal}}$ \\
         & (h:m:s) & ($^{\circ}:^\prime:^{\prime\prime}$) & (\kms) & (pc) & (\kms) & (\kms) & ($10^4$~K~km~s$^{-1}$~pc$^{-2}$) & ($10^5$~M$_\odot$) & (K) & (pc) \\
      \hline
    
      1 & 1:34:18.49 & -29:25:08.16 & 1267.1 & 44.41 $\pm$ 3.54 & 11.44 $\pm$ 1.01 & 9.16 $\pm$ 1.12 & 53.035 $\pm$ 3.525 & 23.335 $\pm$ 1.551 & 9.5 & 357 \\ 
      2 & 1:34:18.45 & -29:25:07.07 & 1282.3 & - & 5.87 $\pm$ 6.31 & 2.38 $\pm$ 7.14 & 16.063 $\pm$ 33.692 & 7.068 $\pm$ 14.825 & 11.2 & 305 \\ 
      3 & 1:34:18.47 & -29:25:06.99 & 1293.4 & 38.21 $\pm$ 2.34 & 12.11 $\pm$ 0.74 & 8.72 $\pm$ 0.61 & 93.500 $\pm$ 3.926 & 41.140 $\pm$ 1.728 & 13.2 & 328 \\ 
      4 & 1:34:18.51 & -29:25:08.43 & 1310.5 & 55.94 $\pm$ 1.57 & 20.15 $\pm$ 0.52 & 17.95 $\pm$ 0.51 & 416.533 $\pm$ 6.462 & 183.274 $\pm$ 2.843 & 17.7 & 385 \\ 
      5 & 1:34:18.65 & -29:25:07.95 & 1304.3 & - & 4.82 $\pm$ 3.60 & 3.62 $\pm$ 5.89 & 4.883 $\pm$ 5.776 & 2.149 $\pm$ 2.541 & 5.6 & 539 \\ 
      6 & 1:34:18.55 & -29:25:07.80 & 1308.3 & - & 4.05 $\pm$ 4.82 & 0.97 $\pm$ 2.38 & 6.225 $\pm$ 12.051 & 2.739 $\pm$ 5.302 & 7.3 & 420 \\
      -&-&-&-&-&-&-&-&-&-&-\\
      356 & 1:34:17.91 & -29:25:07.39 & 1668.8 & - & 4.36$\pm$2.16 & 2.78$\pm$1.79 & 4.461$\pm$2.141 & 1.963$\pm$0.942 & 7.8 & 384 \\
      \hline
    \end{tabular}
    }
\end{table*}

\begin{figure*}
  \centering
  \includegraphics[width=2\columnwidth]{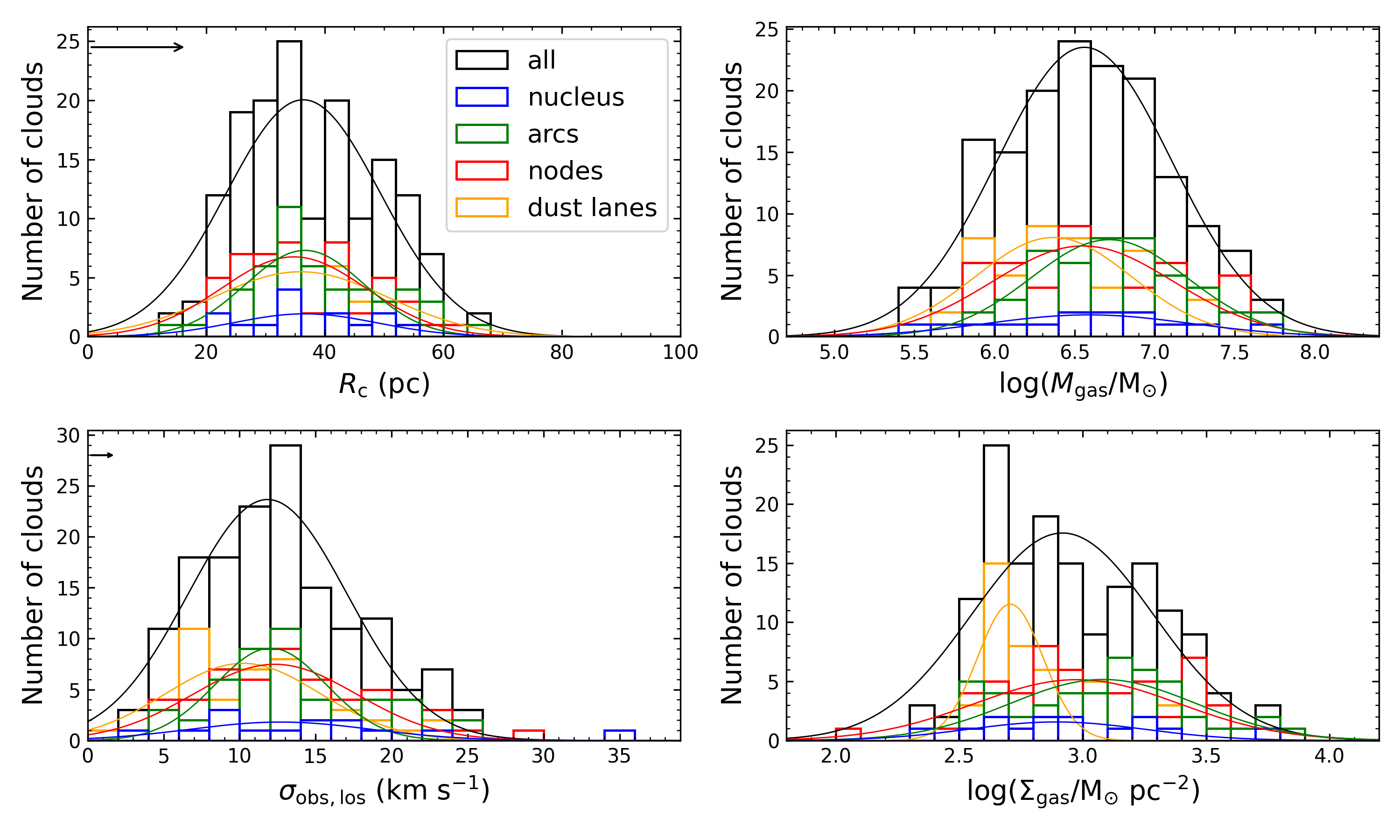}
  \caption{Number distributions of \Rc, $\log(M_{\mathrm{gas}}/\mathrm{M}_{\odot})$, \sigobs\ and $\log(\Sigma_{\mathrm{gas}}/\mathrm{M_\odot~pc^{-2}})$ with Gaussian fits overlaid for the $158$ resolved clouds of NGC~613 (black lines and histograms), and for the clouds in the nucleus (blue), arcs (green), nodes (red) and dust lanes (yellow) only. The black arrows in the two left panels represent our ability to resolve clouds spatially (top-left; $0.43\eta\sqrt{\theta_{\mathrm{maj}}\theta_{\mathrm{min}}}$, where $\theta_{\mathrm{maj}}$ and $\theta_{\mathrm{min}}$ are the synthesised beam's FWHM) and spectrally (bottom-left; channel width of $2$~\kms).}
  \label{fig:histogram}
\end{figure*}

\autoref{fig:histogram} shows the number distributions of \Rc, \Mgas, \sigobs\ and gas mass surface density ($\Sigma_{\mathrm{gas}}\equiv M_{\mathrm{gas}}/\pi R_{\mathrm{c}}^2$) of the resolved clouds of NGC~613. In each panel, the black histogram and overlaid black Gaussian fit show the full cloud sample, while the colour-coded histograms and Gaussians show only the clouds in the nucleus (blue), arcs (green), nodes (red) and dust lanes (yellow), respectively. The means of all the quantities are listed in \autoref{tab:ngc613_gmc}. 

\begin{table}
  \caption{Mean quantities of resolved GMCs.}
  \label{tab:ngc613_gmc}
  \resizebox{\columnwidth}{!}{
    \begin{tabular}{lccccc}
      \hline
      Sample  & \Rc & \Mgas & \sigobs & \surfgas \\
      ($N_{\mathrm{clouds}}$) & (pc) & ($10^{6}$ M$_{\odot}$) & (\kms) & ($10^{3}$ M$_{\odot}$~pc$^{-2}$) \\
      \hline
      All ($158$) & $37.8\pm0.9$ & $7.1\pm0.7$ & $12.8\pm0.4$ & $1.25\pm0.09$ \\
      Nucleus ($11$) & $38.3\pm2.8$ &  $7.7\pm2.8$ & $14.4\pm2.1$ & $1.29\pm0.35$ \\
      Arcs ($48$) & $39.1\pm1.6$ & $9.3\pm1.6$ & $13.5\pm0.7$ & $1.63\pm0.20$ \\
      Nodes ($51$) & $37.2\pm1.6$ & $7.3\pm1.3$ & $13.1\pm0.8$ & $1.29\pm0.14$ \\
      Dust lanes ($48$) & $37.0\pm1.7$ & $4.1\pm0.6$ & $11.2\pm0.7$ & $0.78\pm0.07$ \\
      \hline
    \end{tabular}
}
\end{table}

The resolved clouds of NGC~613 have sizes (\Rc) ranging from $15$ to $70$~pc (top-left panel of \autoref{fig:histogram}) with a mean of $\approx 38$~pc. They have gas masses (\Mgas) ranging from $2.9\times10^5$ to $5.0\times10^7$~\Msun\ (top-right panel of \autoref{fig:histogram}) with a mean of $\approx 7\times10^6$~\Msun. Most ($134/158$) of the resolved clouds are massive ($M_{\mathrm{gas}}\geq10^6$~\Msun). The observed velocity dispersions of the resolved clouds range from $2$ to $36$~\kms\ (bottom-left panel of \autoref{fig:histogram}) with a mean of $\approx 13$~\kms, while the gas mass surface densities range from $100$ to $6300$~\Msun~pc$^{-2}$ (bottom-right panel of \autoref{fig:histogram}) with a mean of $\approx 1200$~\Msun~pc$^{-2}$.

The clouds in the arcs ($9.3\pm1.6\times10^{6}$~\Msun) and the dust lanes ($4.1\pm0.6\times10^{6}$~\Msun) tend to be more and less massive than the clouds in the nucleus ($7.7\pm2.8\times10^{6}$~\Msun) and the nodes ($7.3\pm1.3\times10^{6}$~\Msun), respectively. The clouds in the dust lanes tend to be less turbulent than those in the other three regions ($11.2\pm0.7$~\kms\ vs.\ $14.4\pm2.1$, $13.5\pm0.7$ and $13.1\pm0.8$~\kms). However, there is no significant size variation across the four regions, resulting in higher and lower gas mass surface densities for the clouds in the arcs ($1630\pm200$~\Msun~pc$^{-2}$) and the dust lanes ($780\pm70$~\Msun~pc$^{-2}$) compared to those in the nucleus ($1290\pm350$~\Msun~pc$^{-2}$) and the nodes ($1290\pm140$~\Msun~pc$^{-2}$), respectively. In particular, the clouds in the dust lanes have smaller gas mass surface densities than those in the other three regions (see the double-peaked black histogram in the bottom-right panel of \autoref{fig:histogram}). This implies that the clouds get denser as they migrate into the nuclear ring.

{In \autoref{fig:cloud_compare}, we compare the properties of the GMCs of the MWd, MW central molecular zone (CMZ) and other external galaxies with those of the NGC~613 GMCs using violin plots. The resolved clouds of NGC~613 have comparable sizes but larger masses, velocity dispersions and gas mass surface densities than the clouds of the MWd ($\leq20$~pc spatial resolution; \citealt{heyer2009, rice2016mw, heyer2015_review}; and references therein). On the other hand, they have larger sizes and masses but smaller velocity dispersions than the clouds of the CMZ ($\leq1.5$~pc resolution; \citealt{oka1998, oka2001}).

\begin{figure*}
  \includegraphics[width=2\columnwidth]{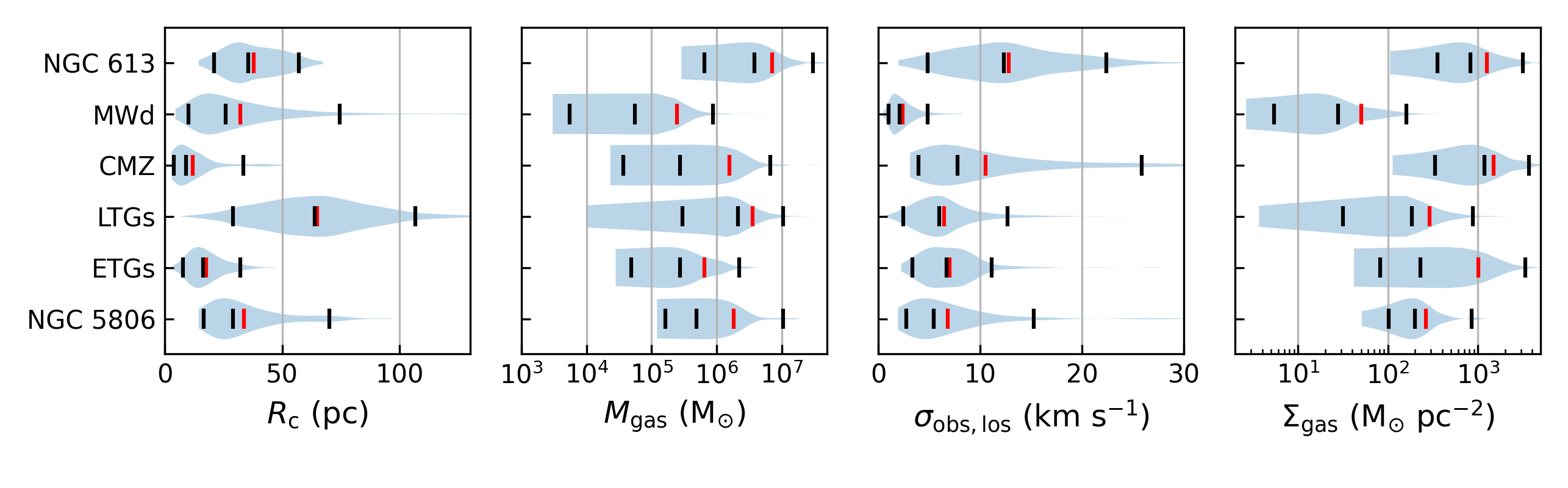}
  \caption{Violin plots showing the distributions of \Rc, $M_{\mathrm{gas}}$, \sigobs\ and $\Sigma_{\mathrm{gas}}$ for the NGC~613 resolved clouds and other extragalactic GMC populations. The ranges for NGC~613 have been derived in this work, while those for the other objects are from the literature: MWd (\citealt{heyer2009,rice2016mw,heyer2015_review}; and references therein), CMZ \citep[][]{oka2001}, LTGs (\citealt{donovan2012, gratier2012M33,colombo2014m51,rosolowsky2021}; Liu et al.\ submitted), ETGs \citep{utomo2015,liu2021ngc4429} and NGC~5806 \citep{choi2023}. The black vertical lines in each violin plot represent the $5^{\rm th}$, $50^{\rm th}$ and $95^{\rm th}$ percentiles of the distribution, while the red line indicates the mean.}
  \label{fig:cloud_compare}
\end{figure*}

Most clouds in LTGs have larger sizes and comparable gas masses but smaller observed velocity dispersions and gas mass surface densities ($10$ -- $80$~pc resolution; e.g.\ \citealt{donovan2012, gratier2012M33, colombo2014m51,rosolowsky2021}; Liu et al.\ submitted) than the NGC~613 clouds. Clouds in ETGs (NGC~4526 and NGC~4429) have smaller sizes, gas masses and observed velocity dispersions ($\leq20$~pc resolution; \citealt{utomo2015, liu2021ngc4429}) than the clouds of NGC~613. The gas mass surface densities of the clouds of NGC~4526 and NGC~4429 are comparable to ($\langle\Sigma_{\mathrm{gas}}\rangle\approx1000$~\Msun~pc$^{-2}$) and smaller than ($\langle\Sigma_{\mathrm{gas}}\rangle\approx150$~\Msun~pc$^{-2}$) that of the clouds of NGC~613, respectively.

The clouds in the morphologically-similar galaxy NGC~5806 ($\approx 24$~pc resolution; \citealt{choi2023}) have very similar sizes ($15$ -- $85$~pc, $\langle R_{\mathrm{c}}\rangle=34.5$~pc), but their gas masses ($1.2\times10^{5}$ -- $3.6\times10^{7}$~\Msun, $\langle M_{\mathrm{gas}}\rangle=6.9\times10^5$~\Msun), velocity dispersions ($1.6$ -- $20$~\kms, $\langle\sigma_{\mathrm{obs,los}}\rangle=6.6$~\kms) and gas mass surface densities ($\langle\Sigma_{\mathrm{gas}}\rangle\approx200$~\Msun~pc$^{-2}$) are much smaller than those of the clouds of NGC~613. For both NGC~613 and NGC~5806, the clouds in the arcs and nodes have gas mass surface densities higher than those in the dust lanes. 

Overall, there is no significant cloud size variation across the four regions of NGC~613. The clouds in the arcs are generally more massive and have larger gas mass surface densities than the clouds in the dust lanes, while the clouds in the nucleus and nodes are in between. The clouds in the dust lanes tend to be less turbulent than those in other the regions.

\subsubsection{Cloud kinematics}

By comparing the observed velocity dispersions (\sigobs) with the gradient-subtracted velocity dispersions (\siggs), we can assess whether bulk motions (due to e.g.\ galaxy rotation and/or shear) contribute significantly to the observed velocity dispersions of the clouds. If the gradient-subtracted velocity dispersions are much smaller than the observed velocity dispersions, bulk motions are dominant in the clouds. \autoref{fig:sigma_compare} shows the relation between \sigobs\ and \siggs. More than $80\%$ of the clouds ($138/158$) of NGC~613 have a small difference between the two velocity dispersions (i.e.\ a ratio between the two velocity dispersions $\sigma_{\mathrm{gs,los}}/\sigma_{\mathrm{obs,los}}>0.7$, with a mean ratio of $0.82$), suggesting that the observed velocity dispersions are dominated by internal turbulent motions rather than bulk motions inherited from larger-scale galaxy rotation. This is similar to the case of NGC~5806, for which more than $60\%$ of the clouds have a ratio larger than $0.7$ \citep{choi2023}.

\begin{figure}
  \centering
  \includegraphics[width=0.9\columnwidth]{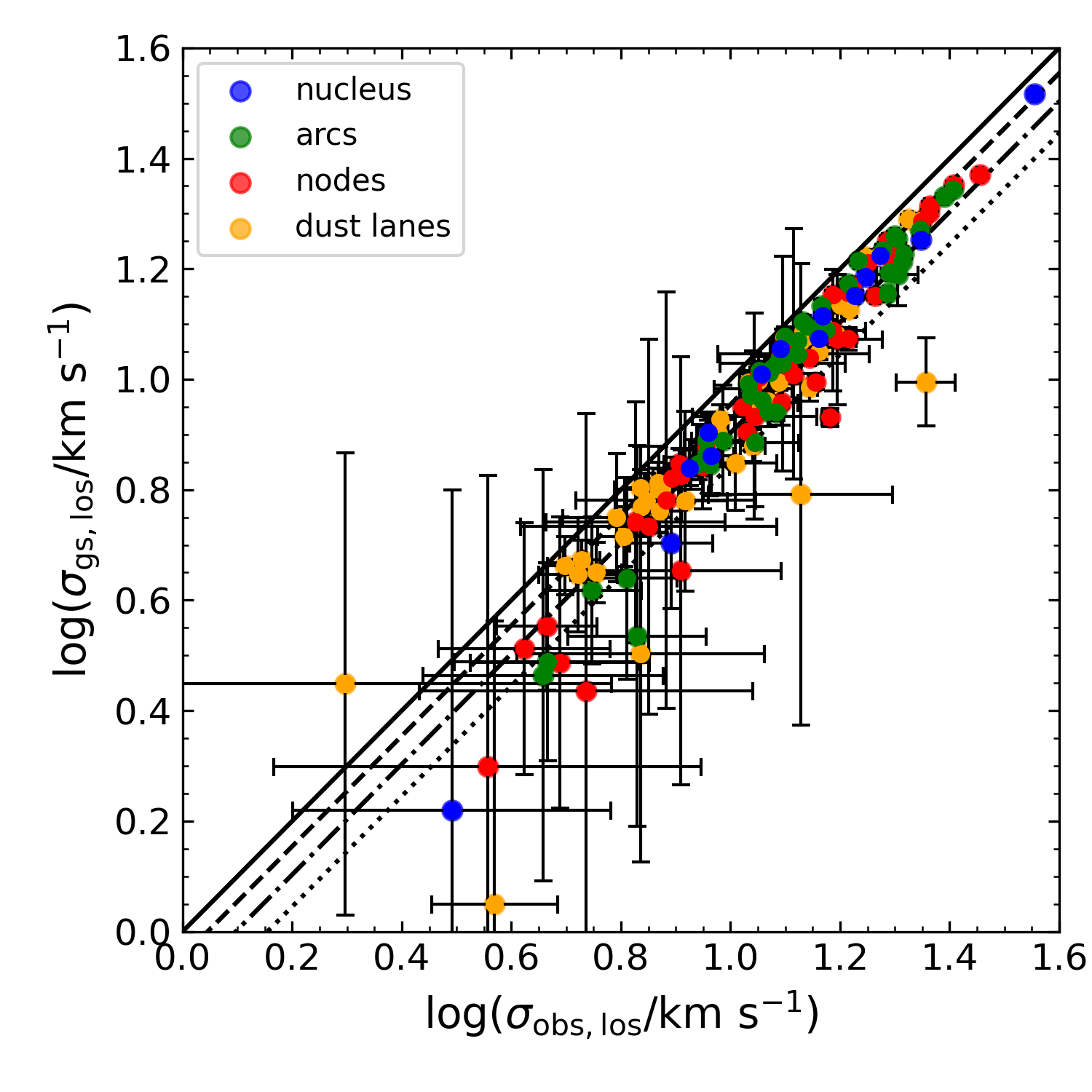}
  \caption{Comparison of the observed (\sigobs) and gradient-subtracted (\siggs) velocity dispersion measures for the $158$ resolved clouds of NGC~613. The four black diagonal lines represent the $1:1$, $1:0.9$, $1:0.8$ and $1:0.7$ ratio, respectively.}
  \label{fig:sigma_compare}
\end{figure}

We also confirm that the rotation axes of the NGC~613 clouds are not well aligned with the isovelocity contours, which is consistent with the case of NGC~5806, supporting that the galaxy rotation does not affect the internal rotation of the clouds. This is similar to the case of the MW \citep{koda2006}, M~31 \citep{rosolowsky2007}, NGC~5806 \citep{choi2023} and NGC~5064 \citep{liu_2023}, but different from that of the ETGs NGC~4526 \citep{utomo2015} and NGC~4429 \citep{liu2021ngc4429}, where the rotation axes are well aligned with the isovelocity contours.

\section{Dynamical states of clouds}
\label{sec:dynamics}

Using the cloud properties calculated above, we now attempt to diagnose the dynamical states of the clouds. A standard relation to characterise clouds is the size -- linewidth relation, known to have the form of a power law (e.g.\ $\sigma_{\mathrm{obs,los}}\propto R_{\mathrm{c}}^{0.5}$, \citealt{larson1981}; \citealt{solomon1987}). Another relation quantifying the gravitational boundness of clouds is that between the virial mass and gas mass, the ratio of which corresponds to the virial parameter of the clouds. We now discuss both relations in turn.

\subsection{Steep size -- linewidth relation}
\label{sec:sizelinewidth}

The size -- linewidth relation is usually interpreted as being due to the internal turbulent motions of clouds \citep[][]{falgarone1991, heyer2004,elmegreen2004_turbulence,Lequeux2005}. In the case of incompressible turbulence, $\sigma_{\mathrm{obs,los}}\propto R_{\mathrm{c}}^{1/3}$ is expected, which comes from the assumption that the energy cascades conservatively down to smaller scales (i.e. constant energy transfer rate, \citealt{kolmogorov1941}). In the case of compressible turbulence, the energy transfer down to a smaller scale is not conservative since the energy can be directly dissipated at large scales via shocks and/or gas compression, leading to steeper size -- linewidth relations with power-law indices of $1/2$ -- $3/5$ (e.g.\ \citealt{cen2021}, \citealt{mckee_ostriker2007}, and references therein).

The left panel of \autoref{fig:larson_only} shows the size -- linewidth relation of all resolved clouds of NGC~613 (coloured circles). There is a clear correlation between size and linewidth, with a Spearman rank correlation coefficient of $0.60$ and a $p$-value of $10^{-16}$. To calculate the best-fitting power-law relation taking into account both \Rc\ and \sigobs\ uncertainties, and to be consistent with previous GMC studies \citep[e.g.][]{liu2021ngc4429, choi2023}, we use a hierarchical Bayesian method called \textsc{Linmix} \citep{kelly2007_linmix}. This yields a best-fitting relation
\begin{equation}
  \log\left(\frac{\sigma_{\mathrm{obs,los}}}{\mathrm{km~s^{-1}}}\right)=(0.77\pm0.11)\log\left(\frac{R_{\mathrm{c}}}{\mathrm{pc}}\right)-(0.10\pm0.18)\,\,\,,
\end{equation}
shown in \autoref{fig:larson_only} as a black solid line. The best-fitting relations of MWd \citep{solomon1987} and CMZ \citep[e.g.][]{kauffmann2017} clouds are also shown as black dashed and black dotted lines, respectively.

\begin{figure*}
  \includegraphics[width=1\columnwidth]{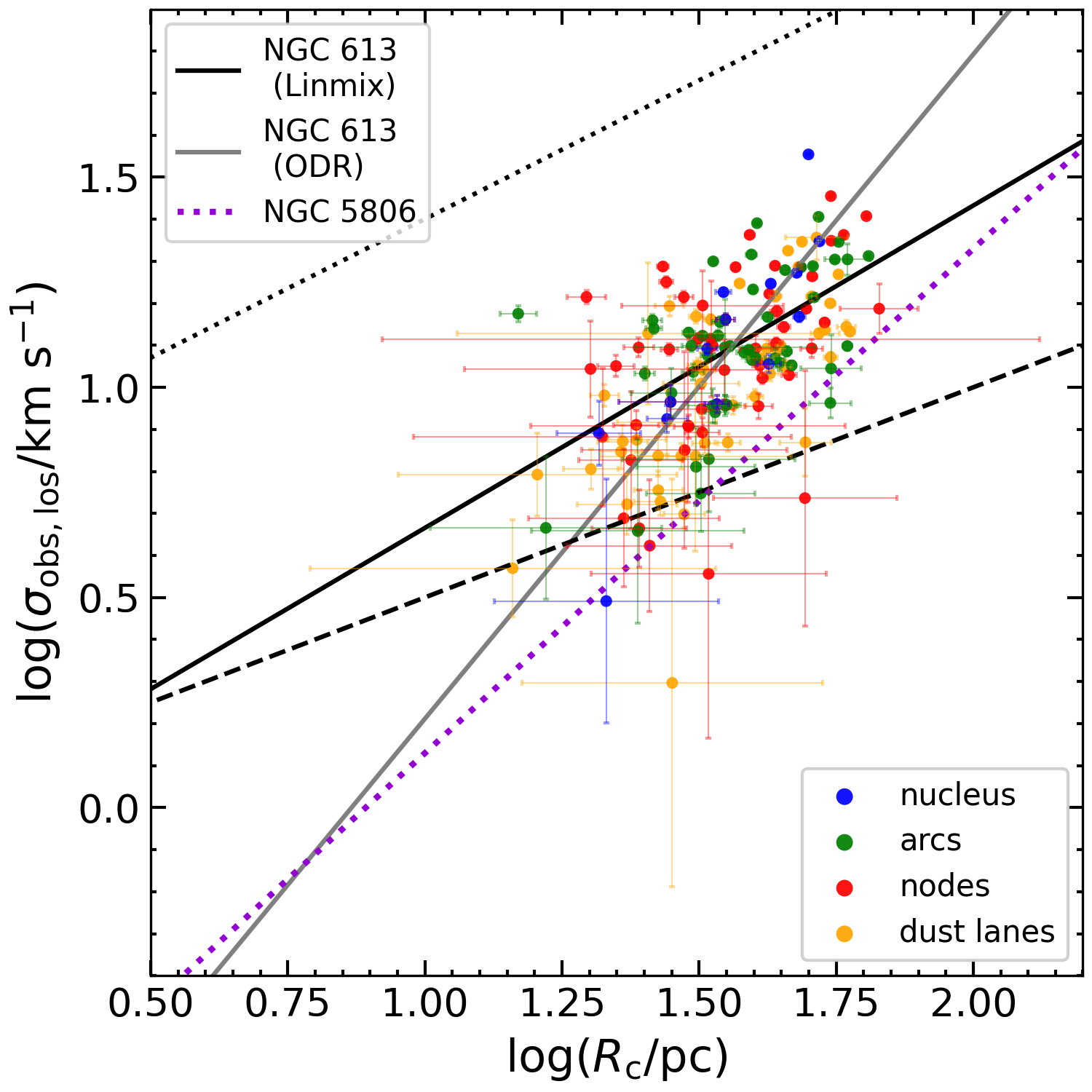}
  \includegraphics[width=1\columnwidth]{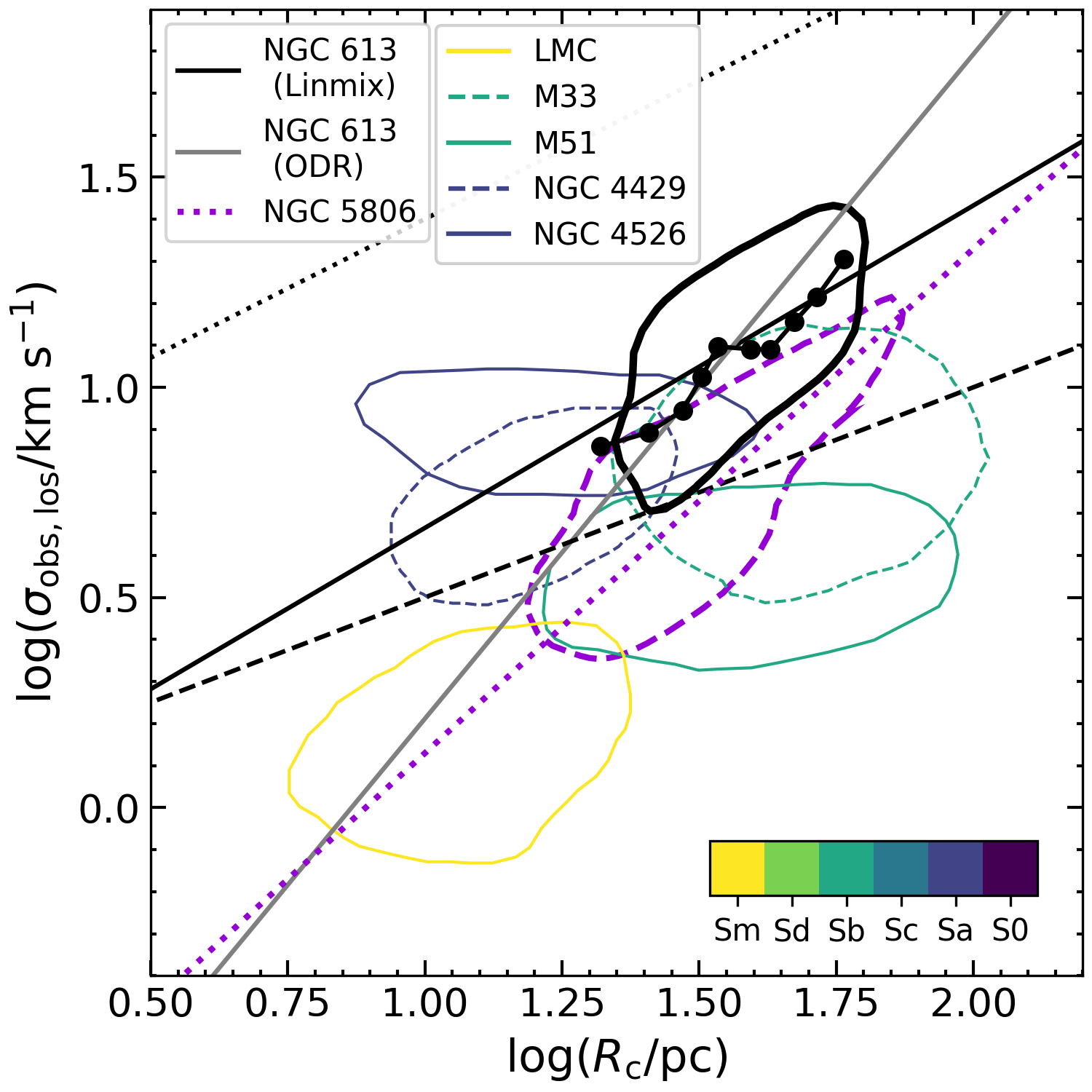}
  \caption{\textbf{Left:} Size -- linewidth relation of the resolved clouds of the nucleus (blue), arcs (green), nodes (red) and dust lanes (yellow) of NGC~613. The black and grey solid lines show the best-fitting power-law relations of all resolved clouds of NGC~613 using the \textsc{Linmix} and ODR algorithm, respectively. The black dashed and black dotted lines show the best-fitting relation of the MWd \citep{solomon1987} and CMZ \citep{kauffmann2017} clouds, respectively. The purple dotted line shows the best-fitting relation of the NGC~5806 clouds. \textbf{Right:} Same as the left panel, but with coloured contours encompassing $68\%$ of the distribution of the data points of a number of galaxies (LMC, \citealt{wong2011LMC}; M~33, \citealt{gratier2012M33}; M~51, \citealt{colombo2014m51}; NGC~4429, \citealt{liu2021ngc4429}; NGC~4526, \citealt{utomo2015}). The contours are colour-coded according to each galaxy' morphological type (see the colour bar). The black and purple contours show the corresponding distributions of NGC~613 and NGC~5806 \citep{choi2023}. The black curve with filled black circles shows the binned relation of NGC~613 (medians in abscissa bins containing equal numbers of clouds).}
  \label{fig:larson_only}
\end{figure*}

The best-fitting slope ($0.77\pm0.11$) is steeper than that of the clouds in the MWd ($0.5\pm0.05$; \citealt{solomon1987}) but is comparable to that of the clouds in the CMZ ($0.66\pm0.18$; \citealt{kauffmann2017}) within the uncertainties, and the zero-point ($0.79\pm0.33$~\kms) is comparable to that of the MWd clouds ($1.0\pm0.1$~\kms; \citealt{solomon1987}) but is much smaller than that of the CMZ clouds ($5.5\pm1.0$~\kms; \citealt{kauffmann2017}). The slope is also steeper than that of the clouds in most nearby galaxies ($0.4$ -- $0.8$; \citealt{rosolowsky2003, rosolowsky2007, bolatto2008, wong2011LMC}) and the predicted slopes of some theoretical models ($1/3$ -- $3/5$; \citealt{kolmogorov1941, cen2021}).%
The grey solid line in \autoref{fig:larson_only}, which shows an even steeper slope, is that from the best-fitting relation derived using the orthogonal distance regression (ODR) algorithm ($\sigma_{\mathrm{obs,los}}\propto R_{\mathrm{c}}^{1.58\pm0.15}$). This slope is even steeper than that of NGC~5806 ($1.20\pm0.10$) and any other galaxy. We however again note that it is the slope yielded by the \textsc{Linmix} algorithm that will be used for the following discussion, to be consistent with previous GMC studies.

The right panel of \autoref{fig:larson_only} shows the size -- linewidth relations of the two ETGs NGC~4526 ($\approx20$~pc spatial resolution and $\approx0.7$~K sensitivity) and NGC~4429 ($\approx13$~pc and $\approx0.5$~K) and the LMC ($\approx11$~pc and $\approx0.3$~K), that have observations with comparable spatial resolutions and sensitivities. While the observations of M~51 \citep{colombo2014m51} and M~33 \citep{gratier2012M33} are more different in term of the spatial resolution, we also include them in the right panel of \autoref{fig:larson_only} to illustrate the size -- linewidth relations of a broader variety of galaxy morphological types. Each coloured contour encompasses $68\%$ of the distribution of the clouds of the associated galaxy, including NGC~5806 (purple contour) and NGC~613 (black contour). The black curve with filled black circles shows the binned trend of NGC~613 from this work. It connects the median velocity dispersions measured in ten bins along the abscissa, each containing an equal number of clouds. Except the LMC ($0.80\pm0.05$, \citealt{wong2011LMC}) and NGC~5806 (purple dotted line), none of these galaxies shows a steeper slope than NGC~613 (M33, $0.45\pm0.02$, \citealt{rosolowsky2003}; M51, no correlation, \citealt{colombo2014m51}; NGC~4429, $0.82\pm0.13$, \citealt{liu2021ngc4429}; NGC~4526, no correlation, \citealt{utomo2015}). The LMC has a comparable slope and a large-scale bar, but comparison with it is not fully justified as the bar is off centred and the whole galaxy is strongly affected by a tidal interaction \citep[e.g.][]{deVaucouleurs1972_lmc_bar, vanderMarel2001_LMC_bar}. In addition, the slope of the NGC~4429 clouds is much shallower ($\approx0.24$; \citealt{liu2021ngc4429}) than that of the NGC~613 clouds once the influence of the large-scale galaxy rotation on the cloud velocity dispersions is removed.

More recently, Physics at High Angular resolution in Nearby GalaxieS (PHANGS) studies (e.g.\ \citealt{sun2020_bar_phangs}, \citealt{sun2022_phangs_bar} and references therein) have been systematically probing the properties of extragalactic molecular gas at intermediate spatial scales. While most of the PHANGS studies were carried out using a beam-by-beam or pixel-by-pixel approach, \citet{rosolowsky2021} identified GMCs in ten nearby galaxies. The spatial resolutions mostly range from $60$ to $90$~pc, with only two galaxies observed with a spatial resolution comparable to that of our study, but \citet{rosolowsky2021} report slopes similar to that of the MWd clouds. The GMC velocity dispersions are however observed to be higher at a given cloud size in the central regions than in the outer regions, potentially also resulting in a steeper slope.

\subsubsection{Size -- linewidth relation in galaxy centres}

We now investigate whether the steep size -- linewidth relation of the NGC~613 clouds is primarily due to the fact that the relation has been measured in the central region of the galaxy. For instance, the large velocity dispersions and relatively steep slope of the CMZ clouds are the most distinguishing features of their size -- linewidth relation, compared to that of the MWd clouds. Moreover, the clouds in the central regions of both NGC~613 and NGC~5806 have steep size -- linewidth relations. On the other hand, the clouds in the centres of other galaxies have size -- linewidth relations much shallower than (or comparable to) that of the MWd clouds (e.g.\ no correlation in NGC~6946, \citealt{wu2017_ngc6946}; slope of $0.6\pm0.1$ in NGC~5064, Liu et al.\ submitted; no correlation in NGC~4526, \citealt{utomo2015}; slope of  $0.29\pm0.11$ in NGC~1387, Liang et al.\ in prep.). Although the clouds in the centre of M~83 have higher velocity dispersions than those of the MWd clouds and have a steep size -- linewidth relation, the correlation is very weak. And while the clumps in the central region of the dwarf lenticular galaxy NGC~404 also have a steep size -- linewidth relation \citep{liu2022_ngc404}, the comparison is arguably inappropriate as the clumps are much smaller ($\approx3$~pc) than the clouds considered in other studies. Therefore, while the centres of galaxies may partially contribute to steep size -- linewidth relations, they are not likely to be the primary driver of them.

\subsubsection{Large-scale bars and nuclear rings}

Although the GMCs of barred-spiral galaxies have been investigated previously (e.g.\ M~83 and NGC~1300; \citealt{hirota2018_m83_bar, maeda2020a_n1300_bar}), only the GMCs in the central region of NGC~5806 display a very steep size -- linewidth relation, with a slope of $\approx 1.2$ (purple dotted line in \autoref{fig:larson_only}). However, among these three galaxies, only the observations of NGC~5806 have a spatial resolution sufficient to resolve GMCs and cover the whole central region. Thus, in \citet{choi2023}, we first suggested that bar-driven gas shocks and inflows toward the nuclear ring, which can potentially cause additional turbulence within the clouds, may be the main reasons for the steep size -- linewidth relation of NGC~5806. Intriguingly, in this work, we find another example of a steep size -- linewidth relation in and around the nuclear ring of a barred spiral galaxy (NGC~613). We thus discuss briefly below whether bar-driven inflows can contribute to the enhanced cloud linewidths and a steep cloud size -- linewidth relation in the nuclear ring of NGC~613. 

Adopting the gas inflow velocities ($V_{\mathrm{in,east}}\approx70$~\kms\ and $V_{\mathrm{in,west}}\approx170$~\kms) inferred by \citet{sato2021}, the total mass inflow rate along each of the two dust lanes can be estimated as 
\begin{equation}
  \label{eq:mass_inflow_rate}
  \begin{split}
    \dot{M}_{\mathrm{in,east}} =\langle\Sigma_{\mathrm{gas}}\rangle W_{\mathrm{in}}V_{\mathrm{in,east}}\approx5.8~{\mathrm{M_\odot~yr^{-1}}}\,\,\,,\\
    \dot{M}_{\mathrm{in,west}} =\langle\Sigma_{\mathrm{gas}}\rangle W_{\mathrm{in}}V_{\mathrm{in,west}}\approx14~{\mathrm{M_\odot~yr^{-1}}}\,\,\,,\\
  \end{split}
\end{equation}
where $\langle\Sigma_{\mathrm{gas}}\rangle\approx400$~\Msun~pc$^{-2}$ and $W_{\mathrm{in}}\approx100$~pc are our measured mean molecular gas mass surface density and width of the gas inflow in both dust lanes, respectively. To examine whether these bar-driven gas inflows carry enough energy to drive turbulence in the nuclear ring, we estimate the total kinetic energy per unit time transported by these gas inflows to the nuclear ring:
\begin{equation}
  \label{eq:energy_inflow_rate}
  \begin{split}
    \dot{E}_{\mathrm{in, total}} & =\dot{E}_{\mathrm{in,east}}+\dot{E}_{\mathrm{in,west}}\\
    & \approx\frac{1}{2}\dot{M}_{\mathrm{in,east}}V_{\mathrm{in,east}}^2+\frac{1}{2}\dot{M}_{\mathrm{in,west}}V_{\mathrm{in,west}}^2\\
    & \approx1.3\times10^5~{\mathrm{M_\odot~km^2~s^{-2}~yr^{-1}}}\,\,\,.\\
    \end{split}
\end{equation}
We note that this is an upper limit, as it assumes that the kinetic energy transported by the inflows is entirely converted into turbulent energy in the nuclear ring. 

To verify whether these bar-driven gas inflows are sufficient to maintain the turbulence in the nuclear ring, we also calculate the turbulent energy dissipation rate $\dot{E}_{\mathrm{diss}}$, defined as 
\begin{equation}
  \label{eq:energy_dissipation_rate}
  \begin{split}
    \dot{E}_{\mathrm{diss}} & \approx M_{\mathrm{NR}}\langle\sigma_{\mathrm{NR}}\rangle^3/(2\,h_{\mathrm{NR}})\\
    & \approx3.4\times10^5~{\mathrm{M_\odot~km^2~s^{-2}~yr^{-1}}}\,\,\,,\\
    \end{split}
\end{equation}
\citep[e.g.][]{maclow_klessen2004}, where $M_{\mathrm{NR}}\approx1.2\times10^9$~M$_\odot$, $\langle\sigma_{\mathrm{NR}}\rangle\approx17$~km~s$^{-1}$ and $h_{\mathrm{NR}}\approx14$~pc are the total mass, mean velocity dispersion and scale height of the molecular gas in the nuclear ring, respectively. This scale height was estimated as $h_{\mathrm{NR}}=\langle\sigma_{\mathrm{NR}}\rangle/\kappa_{\mathrm{NR}}$ \citep{lin_pringle1987}, where $\kappa_{\mathrm{NR}}$ is the epicyclic frequency at the nuclear ring radius, that can be calculated as $\kappa^2_{\mathrm{NR}}\equiv\left.\left(R\frac{d\Omega^2(R)}{dR}+4\Omega^2(R)\right)\right|_{R=R_{\mathrm{NR}}}$, where $\Omega(R)\equiv V_{\mathrm{c}}(R)/R$, $V_{\mathrm{c}}(R)$ is the circular velocity of NGC~613 and $R_{\mathrm{NR}}$ is the radius (at the centre) of the nuclear ring ($R_{\mathrm{NR}}\approx270$~pc). We took $V_{\mathrm{c}}(R)$ from \citet{audibert2019}, who derived a rotation curve from $^{12}$CO(3-2) line observations of NGC~613 assuming the gas is on circular orbits, leading to $\kappa_{\mathrm{NR}}\approx1.2$~km~s$^{-1}$~pc$^{-1}$ and in turn $h_{\mathrm{NR}}\approx14$~pc.

If the kinetic energy input rate is balanced by the turbulent energy dissipation rate (i.e.\ $\dot{E}_{\mathrm{diss}}\approx\dot{E}_{\mathrm{in}}$), the turbulence in the nuclear ring can be sustained by the gas inflows. Although \autoref{eq:energy_inflow_rate} and \autoref{eq:energy_dissipation_rate} contain large uncertainties, the estimated input rate is approximately half of the estimated dissipation rate. This estimation suggests that, while bar-driven molecular gas inflows may be a major contributor to the high velocity dispersions and the steep size -- linewidth relation observed, they are not sufficient to fully explain them.

By comparing the dissipation timescale and the cloud travel time, we can also constrain the necessity of additional turbulence sources. If the dissipation timescale is significantly shorter than the cloud travel time, this suggests that the turbulence of clouds is not maintained during travel from one node to the other and thus clouds are not sustained. The turbulence dissipation timescale of a GMC can be estimated as $t_{\mathrm{diss}}\sim R_{\mathrm{c}}/\sigma_{\mathrm{obs,los}}$ \citep{mckee_ostriker2007}, yielding a range of $1$ -- $9$~Myr and a mean of $3.1$~Myr for the nuclear ring clouds of NGC~613. Taking into account the rotation of the large-scale bar, the cloud travel time between the two nodes can be estimated as
\begin{equation}
  \label{eq:travel_time}
  t_{\mathrm{travel}}=\pi R_{\mathrm{NR}}/(V_{\mathrm{c,NR}}-\Omega_{\mathrm{p}}R_{\mathrm{NR}})
\end{equation}
\citep[][]{koda2021}, where $V_{\mathrm{c,NR}}$ is the circular velocity at the radius of the nuclear ring ($V_{\mathrm{c,NR}}\approx170$~\kms) and $\Omega_{\mathrm{p}}$ is the pattern speed of the large-scale bar. The bar pattern speed of NGC~613 has never been measured, and it is not possible to estimate it from the current observations as they cover only the central region of the galaxy. We can obtain a firm lower limit to the travel time by adopting $\Omega_{\mathrm{p}}=0$ in \autoref{eq:travel_time} (yielding $t_{\mathrm{travel}}\gtrsim4.85$~Myr), but instead we derive the bar pattern speed by assuming that NGC~613 has a flat rotation curve in the outer parts of the disc (i.e.\ around corotation) and by adopting the dimensionless bar rotation rate and projected bar radius measured by \citet{seigar2018_bar}: $\mathcal{R}\equiv R_{\rm CR}/R_{\rm bar}=1.5\pm0.1$ (where $R_{\rm CR}$ is the bar corotation radius and $R_{\rm bar}$ the deprojected bar half-length) and $R_{\rm bar}=90\farcs0\pm4\farcs0$. 
This leads to a bar pattern speed $\Omega_{\mathrm{p}}=V_{\mathrm{c,NR}}/\mathcal{R}R_{\mathrm{bar}}=18.5\pm1.3$~km~s$^{-1}$~kpc$^{-1}$. 
In turn, using \autoref{eq:travel_time}, this leads to a travel time $t_{\mathrm{travel}}\approx5.0$~Myr, so that the average turbulent dissipation timescale is roughly half the travel time between the two nodes and the turbulent dissipation timescales of most nuclear ring clouds ($92/99$) are shorter than $5.0$~Myr. This implies that the turbulence of the clouds would be dissipated before they travel from one node to the other, and therefore that turbulence must be continuously supplied to the nuclear ring clouds (from additional sources) to maintain the high cloud velocity dispersions.

\subsubsection{Other mechanisms inducing the high velocity dispersions of the clouds}  % generating a steep size -- linewidth relation

As discussed in the previous section, in addition to the impact of the bar and the nuclear ring, other mechanisms are needed to explain the excess cloud turbulence in the central region of NGC~613.

\textbf{AGN Feedback}. NGC~613 is known to host an AGN with radio jets \citep[e.g.][]{veron-cetty_1986, hummel1987, miyamoto2017_ngc613}, and AGN feedback is a mechanism known to commonly create high velocity dispersions in molecular gas \citep[e.g.][]{schawinski2009_agn_molecule, simionescu2018_agn_molecule, nesvadba2021_agn_gmc,ruffa2022_agn}. Several simulations have also suggested that AGN feedback can severely affect molecular gas near galaxy centres \citep[e.g.][]{wada2009, mukherjee2018_jet_moleucle}, and have shown that the AGN impact is limited to several hundred parsecs in radius within the galactic discs (but extends beyond $1$~kpc perpendicularly to the discs). In the case of NGC~613, the high \ion{Fe}{ii} to Br$\gamma$ line flux ratio suggests that there is an interaction between the radio jets and the nuclear ring \citep{falcon-barroso2014}, but this appears limited to only a small region of the nuclear ring (see the blue stars in the continuum map of \autoref{fig:moments}). In addition, shocks due to the interaction of the radio jets with the surrounding medium appear to influence only the circumnuclear disc (i.e.\ the nucleus according to our region definition; \citealt{miyamoto2017_ngc613}). \citet{miyamoto2017_ngc613} also reported that there are only two small regions where the shock tracer SiO(2-1) is detected near the nuclear ring, as shown in \autoref{fig:moments}. Therefore, the impact of the AGN feedback on the molecular gas in the nuclear ring and its vicinity appears limited to small regions and thus only a portion of the NGC~613 clouds.

\textbf{Stellar Feedback}. Early stellar feedback can contribute to cloud turbulence in the forms of stellar winds and the expansion of \ion{H}{ii} regions \citep[e.g.][]{maclow_klessen2004}. \citet{mazzuca_nuclear_ring} measured a star-formation rate of $\approx2.2$~M$_{\odot}$~yr$^{-1}$ using the H$\alpha$ emission from the nuclear ring (within a radius of $400$~pc) of NGC~613. This ring was also observed in the Br$\gamma$ line using the Very Large Telescope SINFONI integral-field spectrograph, revealing sequential star-forming clumps with an age gradient along the nuclear ring \citep{boker2008,falcon-barroso2014}. As Br$\gamma$ is predominantly produced by the dense \ion{H}{ii} regions surrounding young OB stars \citep{ho1990_brgamma} and can be used to trace recent star formation \citep{pasha2020_brgamma}, early stellar feedback is likely to affect the molecular gas of the nuclear ring and thus drive the cloud turbulence, suggesting that it can contribute to the high cloud velocity dispersions.
Once a massive star dies, it explodes and thus injects significant energy into its surroundings, so-called supernova (SN) feedback. SN feedback is well known as one of the major energy sources regulating the evolution of the interstellar medium \citep{maclow_klessen2004}. It has in fact been suggested that SN feedback is the dominant mechanism driving the turbulence of molecular gas \citep[e.g.][]{padoan2016_sn_turbulence}. Using near-infrared line diagnostic diagrams \citep{colina2015_nir_test}, \citet{audibert2019} showed that the nuclear ring of NGC~613 has characteristics of young stars and/or aged ($\approx8$ -- $40$~Myr) star-forming clumps dominated by SN. SN can thus also contribute to the high cloud velocity dispersions present in the nuclear ring of NGC~613.

\textbf{Accretion-driven turbulence}. \citet{goldbaum2011_accretion} suggested that gas accretion, i.e.\ the flow of surrounding cold gas onto a molecular cloud, and star formation feedback (in the form of \ion{H}{ii} region expansion) contribute roughly equally to the turbulent kinetic energy over the lifetime of a cloud. However, in high gas mass surface density environments, gas accretion dominates the cloud energy budget (over star-formation feedback). As the centre of NGC~613 has high molecular gas mass surface densities, and there are gas inflows toward the nuclear ring via the large-scale bar, accretion-driven turbulence is likely to contribute to the high velocity dispersions of the clouds. Indeed, the molecular gas mass and surface density of the clouds in the nuclear ring are higher than those of the clouds in the dust lanes, implying there is gas accretion that can induce cloud turbulence in the nuclear ring. We note that accretion-driven turbulence is distinct from turbulence driven by bar-driven gas inflows in terms of both physical scale and location. The former occurs at the cloud scale and can take place at any position in and around the nuclear ring, while the latter indicates an energy transfer from large-scale gas inflows to the nuclear ring molecular gas and can only occur near the nodes.

\textbf{Cloud-cloud collisions}. Cloud-cloud collisions have been proposed as another mechanism that can limit cloud lifetimes and act as a source of turbulence within clouds \citep[e.g.][]{tan2000_ccc, Li2018_ccc, wu2018_collision_turbulence}. For instance, using simulations \citet{wu2018_collision_turbulence} suggested that GMC collisions can create and maintain large turbulence within dense gas structures. To assess whether cloud-cloud collisions are important in the nuclear ring of NGC~613, we calculate the cloud-cloud collision timescale. This can be estimated as $t_{\mathrm{coll}}=1/N_{\mathrm{mc}}D_{\mathrm{c}}\sigma_{\mathrm{cc}}$ \citep{koda2006}, where $N_{\mathrm{mc}}$ is the cloud number surface density, $D_{\mathrm{c}}$ is the mean cloud diameter ($2\,\langle R_{\mathrm{c}}\rangle\approx76$~pc; see \autoref{tab:ngc613_gmc}) and $\sigma_{\mathrm{cc}}$ is the cloud-cloud velocity dispersion, generally assumed to be $\approx10$~\kms\ \citep[e.g.][]{koda2006, inutsuka2015_cloud}. To calculate $N_{\mathrm{mc}}$, we consider the $99$ nuclear ring clouds within an elliptical annulus of inner semi-major axis length $240$~pc, outer semi-major axis length $385$~pc and ellipticity $0.3$, yielding $N_{\mathrm{mc}}\approx470$~kpc$^{-2}$. Overall, this yields $t_{\mathrm{coll}}\approx3.0$~Myr, more than half of the cloud travel time between the nodes (ratio of $\approx0.6$).
%{\bf (MB: Correct for the new cloud travel time.)} 
However, \citet{Li2018_ccc} suggested that cloud-cloud collisions play an important role in driving cloud turbulence only when collisions are more frequent (collision to orbital timescale ratio ranging from $0.1$ to $0.2$). Thus, while cloud-cloud collision can contribute to cloud turbulence, they are not likely to be a significant source of turbulence in the nuclear ring of NGC~613. We note that the collision timescale at the nodes should be shorter than the estimated collision timescale above since the cloud number density is higher at the nodes than at the arcs. Thus, cloud-cloud collisions could be more important at the nodes than the arcs. 

In summary, no potential source of turbulence appears to be dominant over the others in the nuclear ring of NGC~613. Among them, stellar feedback, gas accretion and cloud-cloud collisions are all plausible sources for the elevated velocity dispersions of the clouds and the steep size -- linewidth relation. We note that all the mechanisms suggested above can also destroy clouds depending on their strengths. Thus, a more quantitative investigation of stellar feedback and accretion-driven turbulence is required to better understand their impact on the turbulence of the clouds in NGC~613.

\subsubsection{Possible mechanisms for a steep size -- linewidth relation}

Although we suggested several mechanisms that can enhance the linewidth of the clouds in and around the nuclear ring of NGC~613, it is still unclear why those clouds show a steep size -- linewidth relation. To show a steep size -- linewidth, one of two things (or both) must happen. Either smaller clouds have lower turbulence or larger clouds are selectively injected with more turbulence.

Firstly, the interstellar medium is highly compressible, indicating that turbulent energy may not fully cascade down to small scales, but rather be spent on shocks and/or gas compression \citep[e.g.][]{maclow_klessen2004}. As a consequence, smaller clouds can exhibit considerably lower turbulence than larger ones, resulting in a steep size-linewidth relation. However, various numerical simulations have shown that the slope of the size-linewidth relation for strongly compressible flows typically ranges from $0.25$ to $0.5$ (e.g.\ \citealt{Padoan2006_slope,padoan2007_slope}, \citealt{schmidt2009_slope}, \citealt{mckee_ostriker2007} and references therein) and rarely exceeds this range. In addition, \citet{cen2021} recently suggested a new size -- linewidth relation with a slope of $0.6$ from a theoretical approach, which is still not sufficiently high to explain our findings. \citet{kauffmann2017} also attempted to suggest plausible explanations for an unusually steep size-linewidth relation of CMZ clouds. They suggested that the prevalence of shocks, possibly resulting from cloud-cloud collisions and evidenced by SiO emission, could contribute to elevated velocity dispersions. They also proposed stellar feedback and intermittent activity from Sgr~A$^*$ as potential additional mechanisms contributing to elevated linewidths and a steep size-linewidth relation. Nevertheless, the absence of a dominant mechanism and the lack of theoretical arguments explaining the steep size-linewidth remain unaddressed. Therefore, whether the turbulent energy of molecular gas in and around the nuclear ring cascades down to small scales less efficiently (and the mechanisms that cause this) remains to be clarified.

Another way to create a steep size -- linewidth relation is to inject more turbulence into the larger clouds. All the mechanisms suggested above can increase the velocity dispersions of clouds of all sizes, but each mechanism may have a characteristic injection scale, that can in turn lead to a steep size -- linewidth relation. This is thus another aspect that requires further study.

\subsection{Cloud virialisation}\label{sec:virial}

We assess the dynamical state of a cloud using the virial parameter
\begin{equation}
  \label{eq:virial_parameter}
  \alpha_{\mathrm{obs,vir}} \equiv \frac{2K}{|U|} = \frac{\sigma_\mathrm{obs,los}^2 R_{\rm c}}{b_{\rm s} G M_\mathrm{gas}} \equiv \frac{M_\mathrm{obs,vir}}{M_\mathrm{gas}}.
\end{equation}
Here, $K$ and $U$ are the kinetic and gravitational potential energies, respectively, and $b_{\rm s}$ is a geometrical factor that depends on the internal structure of a cloud, for which we adopt $b_{\rm s} = 1/5$ assuming the clouds are homogeneous and spherical.
Note that we have customarily defined the virial mass $M_\mathrm{obs,vir} \equiv \sigma_\mathrm{obs,los}^2 R_{\rm c}/(b_{\rm s} G)$ in the last equality of \autoref{eq:virial_parameter}.
The virial theorem states that a turbulent pressure-dominated cloud would collapse if $\alpha_\mathrm{obs,vir} < 1$ (or equivalently, $M_\mathrm{gas} > M_\mathrm{obs,vir}$) while it must be either confined by external pressure and/or magnetic fields or disperse when $\alpha_\mathrm{obs,vir} \gtrsim 2$.

The top panel of \autoref{fig:larson_obs} shows the virial masses of the resolved clouds of NGC~613 (calculated using the observed velocity dispersion \sigobs; see \autoref{eq:virial_parameter}) as a function of their gas masses, overlaid with the best-fitting power law (black solid line). The black dashed and black dotted lines indicate $\alpha_{\mathrm{obs,vir}}=1$ and $\alpha_{\mathrm{obs,vir}}=2$, respectively. The best-fitting power law is
\begin{equation}
  \log\left(\frac{M_{\mathrm{obs,vir}}}{\mathrm{M_{\odot}}}\right)=(0.94\pm0.03)\log\left(\frac{M_{\mathrm{gas}}}{\mathrm{M_{\odot}}}\right)+(0.58\pm0.02)\,\,\,,
\end{equation} 
implying that the resolved clouds of NGC~613 are virialised on average. 

\begin{figure}
  \centering
  \includegraphics[width=1\columnwidth]{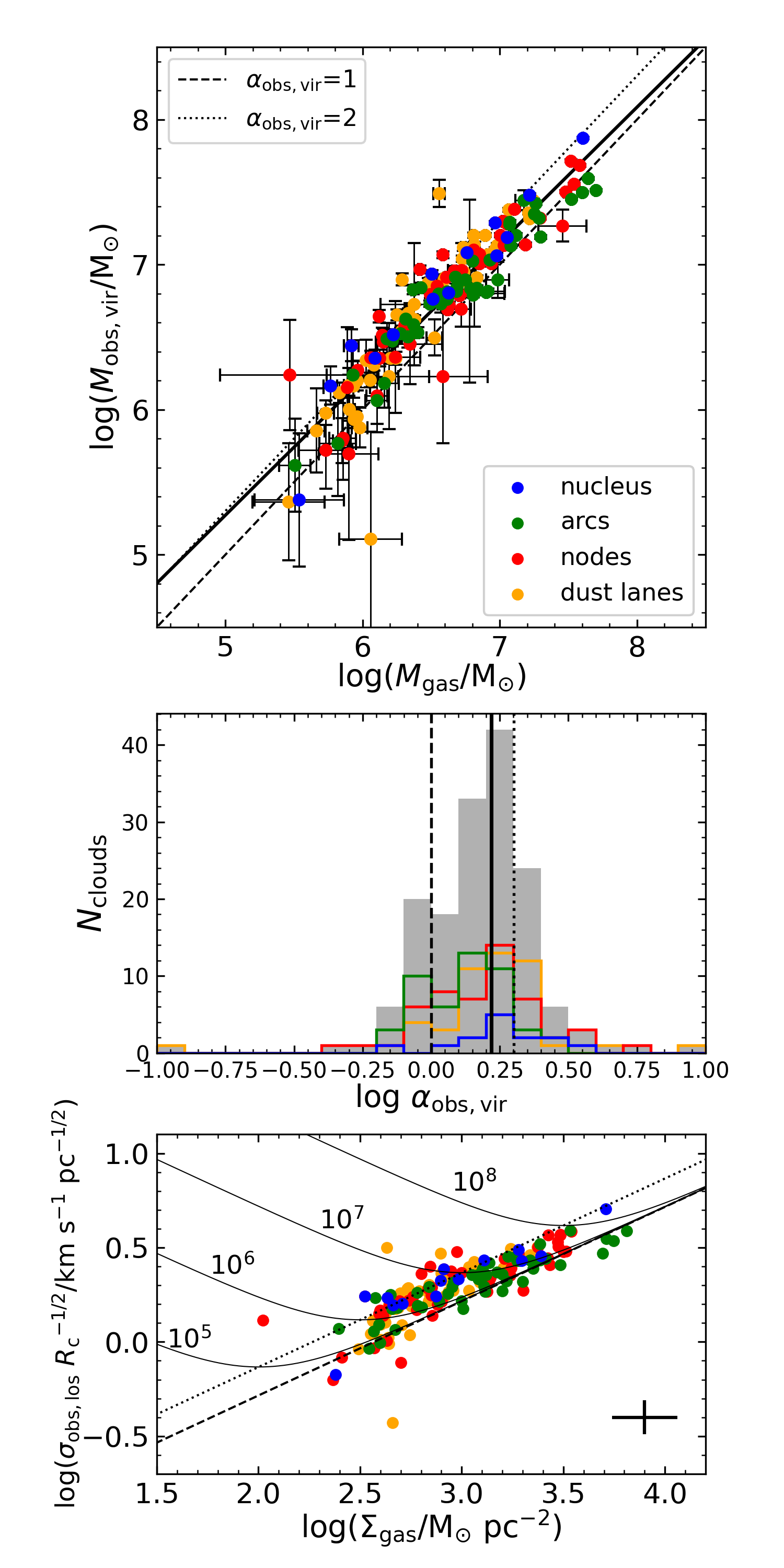}
  \caption{\textbf{Top}: Molecular gas mass -- virial mass relation for the resolved clouds of the nucleus (blue), arcs (green), nodes (red) and dust lanes (yellow) of NGC~613. The black solid line shows the best-fitting power-law relation, while the black dashed and black dotted lines indicate $\alpha_{\mathrm{obs,vir}}=1$ and $\alpha_{\mathrm{obs,vir}}=2$, respectively. \textbf{Middle}: Distribution of $\log\alpha_{\mathrm{obs,vir}}$, where the black solid line shows the mean virial parameter $\langle\alpha_{\mathrm{obs,vir}}\rangle=1.66$, and the black dashed and black dotted lines indicate $\alpha_{\mathrm{obs,vir}}=1$ and $\alpha_{\mathrm{obs,vir}}=2$, respectively. \textbf{Bottom}: Relation between molecular gas mass surface density and $\sigma_{\mathrm{obs,los}}R_{\mathrm{c}}^{-1/2}$ for the same clouds. The black dashed and black dotted diagonal lines show the solutions for a simple (i.e.\ $\alpha_{\mathrm{obs,vir}}=1$) and a marginal (i.e.\ $\alpha_{\mathrm{obs,vir}}=2$) virial equilibrium, respectively. The V-shaped black solid curves show solutions for pressure-bound clouds at different pressures ($P_{\mathrm{ext}}/k_{\mathrm{B}}=10^5$, $10^6$, $10^7$ and $10^8$~K~cm$^{-3}$). A typical uncertainty is shown as a black cross in the bottom right of the panel.}
  \label{fig:larson_obs}
\end{figure}

To investigate the virialisation of the resolved clouds of NGC~613 further, we analyse the distributions of \virialpha\ for all the GMCs of NGC~613 and those in each region individually, as shown in the middle panel of \autoref{fig:larson_obs}. The mean (median) of $\alpha_{\mathrm{obs,vir}}$ is $1.66$ ($1.58$), indicating that on average the clouds are marginally bound (black solid line). However, \virialpha\ has a broad distribution and about $59\%$ of the clouds ($93/158$) lie between $\alpha_{\mathrm{obs,vir}}=1$ and $\alpha_{\mathrm{obs,vir}}=2$. About $23\%$ of the clouds ($36/158$) have $\alpha_{\mathrm{obs,vir}}>2$, while about $18\%$ of the clouds ($29/158$) have $\alpha_{\mathrm{obs,vir}}<1$. Among the four regions, clouds in the arcs tend to have smaller virial parameters (mean of $1.39$) than those of the other three regions (mean of $1.71$ to $1.84$). This indicates that the clouds in the arcs are more strongly gravitationally bound than those in the other three regions. However, we note that the virial parameters scale inversely with the CO-to-molecule conversion factor \Xco. For instance, \citet{sato2021} used $X_{\mathrm{CO}}=0.8\times10^{20}$cm$^{-2}$(K~km~s$^{-1}$)$^{-1}$ and a different cloud finding algorithm (\textsc{Dendrogram}; \citealt{rosolowsky2008_dendrogram}) for NGC~613, yielding a mean cloud virial parameter of $\approx2.8$, approximately $70\%$ larger than ours. It is worth noting that our analysis is more elaborate than theirs, possibly also contributing to the observed differences. Nevertheless, similar to our results, they also reported that the clouds in the arcs tend to have virial parameters lower than those in the dust lanes. Interestingly, low conversion factors have been reported in galaxy centres \citep[e.g.][]{Sandstrom2013}, and \citet{teng2022,teng2023} reported that the conversion factors of barred galaxy centres are $4$ -- $15$ times lower than that of the MWd. These results imply that the virial parameters of the clouds of NGC~613 could be at least four times larger than currently calculated, which would in turn imply that the clouds are not gravitationally bound but either confined by external forces or short-lived.

The bottom panel of \autoref{fig:larson_obs} shows the correlation between molecular gas mass surface density ($\Sigma_{\mathrm{gas}}$) and $\sigma_{\mathrm{obs,los}}R_{\mathrm{c}}^{-1/2}$ for all the resolved clouds of NGC~613, providing another perspective to assess the dynamical states of the clouds \citep{field2011}. Regardless of how well clouds obey the size -- linewidth relation, if the clouds are virialised they should be clustered around $\sigma_{\mathrm{obs,los}}R_{\mathrm{c}}^{-1/2}=\sqrt{\pi\alpha_{\mathrm{obs,vir}}Gb_{\mathrm{s}}\Sigma_{\mathrm{gas}}}$, indicated by the black dashed ($\alpha_{\mathrm{obs,vir}}=1$) and black dotted ($\alpha_{\mathrm{obs,vir}}=2$) diagonal lines. If the clouds are not virialised ($\alpha_{\mathrm{obs,vir}}\gtrsim2$) but are nevertheless long-lived, external pressure ($P_{\mathrm{ext}}$) should be important to confine them (otherwise the clouds are likely transient structures). In this case, the clouds will be distributed around the black solid V-shape curves in the bottom panel of \autoref{fig:larson_obs}:
\begin{equation}
  \label{eq:pressure_virial}
  \sigma_{\mathrm{obs,los}}R_{\mathrm{c}}^{-1/2}=\sqrt{\frac{\pi\alpha_{\mathrm{obs,vir}}G\Sigma_{\mathrm{gas}}}{5}+\frac{4}{3}\frac{P_{\mathrm{ext}}}{\Sigma_{\mathrm{gas}}}}
\end{equation}
\citep{field2011}. 

The molecular gas mass surface densities of the resolved clouds of NGC~613 are broadly distributed, varying by $1.5$ orders of magnitude, and are positively correlated with $\sigma_{\mathrm{obs,los}}R_{\mathrm{c}}^{-1/2}$. Given the uncertainties, some clouds with $\alpha_{\mathrm{obs,vir}}>2$ distributed across the V-shaped curves do seem to be bound by high external pressures ($10^6\lesssim P_{\mathrm{ext}}/k_{\mathrm{B}}\lesssim10^8$~\punit, if indeed they are bound). However, the majority of the clouds are located between the $\alpha_{\mathrm{obs,vir}}=2$ and $\alpha_{\mathrm{obs,vir}}=1$ lines, implying that external pressure is unlikely to be important to constrain them. However, we again note that this is impacted by the uncertainties of the CO-to-molecule conversion factor. For example, if we adopt a conversion factor $4$ times lower, the cloud molecular gas mass surface densities and virial parameters are in turn $4$ times lower. Should that be the case, the clouds of NGC~613 would be unbound (or pressure-bound), similar to the states of the CMZ clouds.

In summary, \autoref{fig:larson_only} shows that the size -- linewidth relation of the resolved clouds of NGC~613 has a slope that is at least as steep as that of CMZ clouds (or much steeper if we adopt the ODR algorithm). For NGC~613, bar-driven gas inflows alone are not sufficient to explain the increased linewidths of the clouds, and other processes such as stellar feedback, cloud-cloud collisions and/or gas accretion must also contribute. Most of the clouds are marginally gravitationally bound ($\langle\alpha_{\mathrm{obs,vir}}\rangle\approx1.7$ with $X_{\rm CO}$ of the MW) and do not seem affected by external pressure (\autoref{fig:larson_obs}).

\subsection{Nuclear ring clouds}
\label{sec:discussion_gmc_ring}

As discussed in the previous section (\autoref{sec:sizelinewidth}), gas inflows driven by the bar are likely to have an impact on the cloud characteristics within the nuclear ring, an effect that has been explored in some previous studies \citep[e.g.][]{salak2016, sato2021, choi2023}. Additionally, some galaxies with a nuclear ring show an azimuthal stellar age gradient within it \citep{mazzuca_nuclear_ring}. Therefore, it is important to investigate whether the nodes have any influence on the cloud properties and whether the cloud properties change with azimuthal angle along the nuclear ring of NGC~613. \autoref{fig:ring_cloud} presents the number of clouds and other cloud properties as a function of azimuthal angle (measured counterclockwise from the western node): virial parameter, gas mass, velocity dispersion, size and gas mass surface density. Interestingly, the number of clouds decreases from one node to the other (see panel~(a) of \autoref{fig:ring_cloud}), yet other cloud properties show no discernible trend and exhibit large scatters as a function of azimuthal angle (see panels~(b) -- (f) of \autoref{fig:ring_cloud}).

\begin{figure*}
\centering
  \includegraphics[width=2\columnwidth]{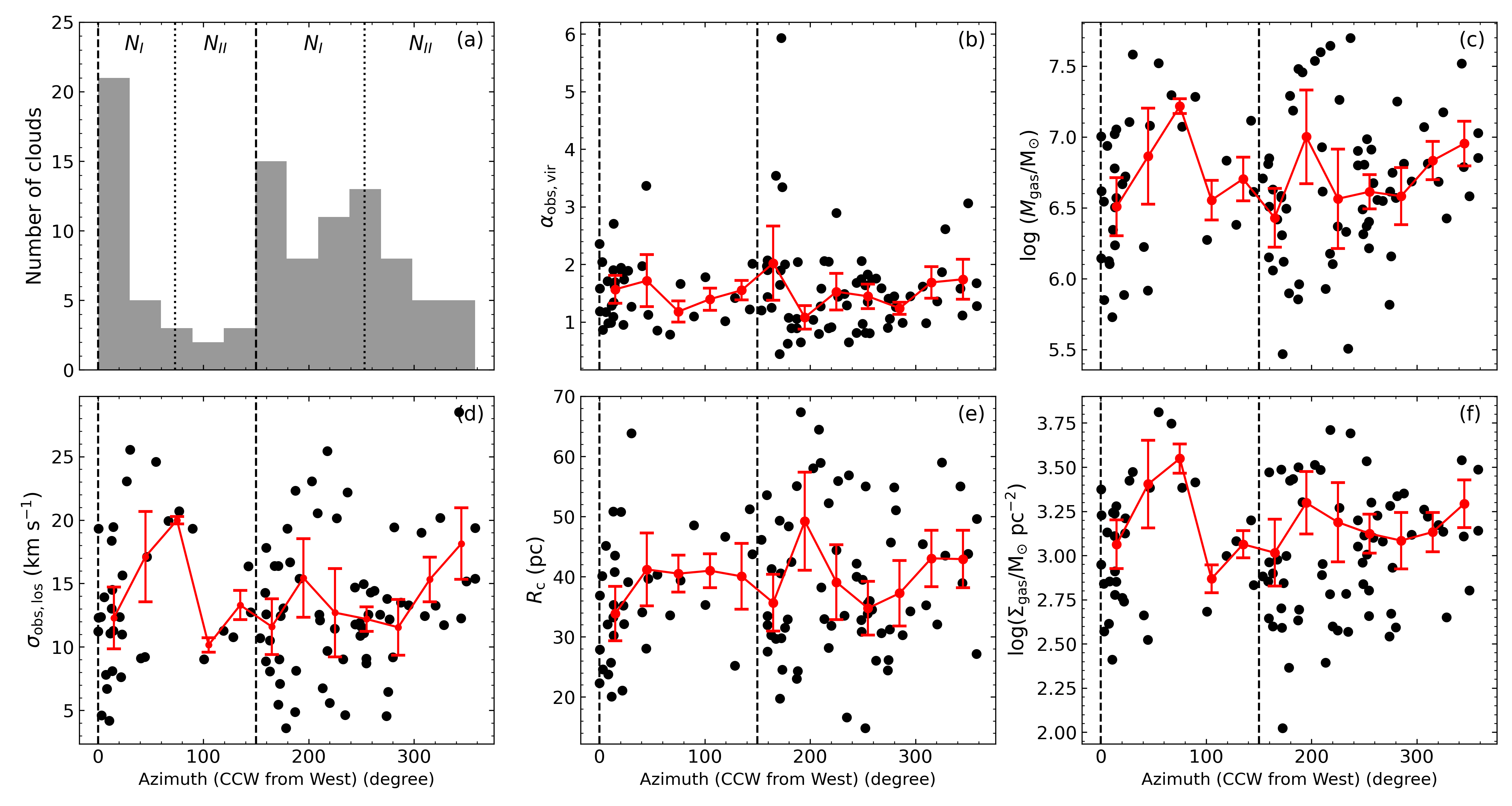}
  \caption{Properties of the clouds in the nuclear ring (both arcs and nodes) of NGC~613, as a function of the azimuthal angle (measured counter-clockwise from the western node). From left to right, top to bottom: number of resolved clouds, virial parameter, molecular gas mass, velocity dispersion, size and molecular gas mass surface density. The red data points are averages in azimuthal bins of width $30\degr$, while the red error bars indicate the $1\sigma$ scatter within each bin. Black vertical dashed lines indicate the positions of the two nodes, and black vertical dotted lines indicate the boundary between two adjacent zones of each half of the nuclear ring (top-left panel only), used to calculate \autoref{eq:Flost}.}
  \label{fig:ring_cloud}
\end{figure*}

In our previous work, we reported an analogous trend in NGC~5806 (see Figure~12 of \citealt{choi2023}), i.e.\ the number of clouds decreases gradually from one node to the other but none of the cloud properties varies significantly along the nuclear ring. We speculated that these trends are due to the gas inflows along the large-scale bar inducing violent collisions at the nodes, leading to the formation of many clouds, which are then gradually destroyed while moving along the nuclear ring. We also suggested that cloud-cloud collisions, shear and stellar feedback are plausible mechanisms for the changes of cloud number and properties along the nuclear ring, as their timescales appear to be shorter than the estimated cloud lifetimes in NGC~5806. It is therefore interesting to verify whether the same reasoning applies to NGC~613.

To compare the different mechanisms, we first calculate the characteristic cloud lifetime, adopting the same method used in our previous work \citep{choi2023}. The azimuthal variation of the number of clouds depends on the characteristic cloud lifetime, that can be estimated as
\begin{equation}\label{eq:lifetime}
t_{\mathrm{lifetime}}=\frac{t_{\mathrm{travel}}}{2}\frac{1}{F_{\mathrm{lost}}}\,\,\,,
\end{equation}
where $F_{\mathrm{lost}}$ is the fraction of clouds lost (i.e.\ the decline of the number of clouds) as they move between the two nodes. This argument is similar to that introduced by \citet{meidt2015} to estimate the cloud lifetimes in the inter-arm region of a spiral galaxy. They suggested that the number changes as the cloud travels and evolves from one region to another, and that the cloud lifetime can be inferred from this. This formalism relies on a number of assumptions: 1) cloud formation is negligible between the two zones and 2) clouds follow circular paths. In the case of NGC~613, we assume that most cloud formation occurs near the nodes. There are non-circular motions, but they are neither significant nor relevant in the nuclear ring. The fraction of clouds lost between the two nodes is
\begin{equation}
  \label{eq:Flost}
  F_{\mathrm{lost}}=\frac{N_{\mathrm{I}}-N_{\mathrm{II}}}{N_{\mathrm{I}}}\,\,\,,
\end{equation}
where $N_{\mathrm{I}}$ and $N_{\mathrm{II}}$ are the number of clouds in two adjacent zones that span equal ranges of azimuth (see panel (a) of \autoref{fig:ring_cloud}). We count the numbers of clouds in the first half of the nuclear ring (from the western node to the eastern node), yielding $N_{\mathrm{I}}=27$ and $N_{\mathrm{II}}=8$ (and thus $F_{\mathrm{lost}}=0.67$), while in the second half (from the eastern node to the western node) this yields $N_{\mathrm{I}}=35$ and $N_{\mathrm{II}}=24$ (and thus $F_{\mathrm{lost}}=0.31$).

Combined with our estimated travel time, these two fractions of lost clouds yield two characteristic cloud lifetime estimates, that we take as a range, $t_{\mathrm{lifetime}}=3.7$ -- $8.1$~Myr. This cloud lifetime is larger than that of the clouds in the CMZ ($1$ -- $4$~Myr; e.g.\ \citealt{kruijssen2015_cmz_lifetime, jeffreson2018_cmz}), but is smaller than that of the clouds in the LMC ($\approx11$~Myr; \citealt{ward2022_lmc_lifetime}), the central $3.5$~kpc radius of M~51 ($20$ -- $50$~Myr; \citealt{meidt2015}), nearby galaxies ($5$ -- $100$~Myr; \citealt{jeffreson2018, chevance2020,kim2022_gmc}), between spiral arms of disc galaxies ($\approx100$~Myr; e.g.\ \citealt{koda2009_gmc_lifetime}) and NGC~5806 ($\approx7.4$~Myr; \citealt{choi2023}).

Prior to making a comparison of the estimated cloud lifetime with other timescales, we calculate the shear timescale, which can also regulate cloud's lifetimes, especially in galaxy centres where strong shear can lead to cloud disruption and/or mass loss \citep[e.g.][]{meidt2015, jeffreson2018}. We estimate the shear timescale as $t_{\mathrm{shear}}=1/2A$ \citep{liu2021ngc4429}, where $A\equiv\frac{1}{2}\left(\frac{V_{\mathrm{c}}(R_{\mathrm{NR}})}{R_{\mathrm{NR}}}-\left.\frac{dV_{\mathrm{c}}}{dR}\right|_{R_{\mathrm{NR}}}\right)\approx0.14$~\kms~pc$^{-1}$ is Oort's constant evaluated at the nuclear ring using the aforementioned rotation curve, yielding $t_{\mathrm{shear}}\approx3.5$~Myr. 

Consequently, these other timescales ($t_{\mathrm{coll}}\approx3$~Myr and $t_{\mathrm{shear}}\approx3.5$~Myr) tend to be shorter than the characteristic cloud lifetime ($t_{\mathrm{lifetime}}=3.7$ -- $8.1$~Myr), indicating that they could also play a role and affect the number of clouds and other cloud properties along the nuclear ring, in addition to the stellar feedback discussed previously. It is, therefore, still unclear which mechanism primarily regulates the cloud lifetimes and causes the number of clouds to decrease with azimuth while other cloud properties remain roughly constant. Quantifying the impact of each of these mechanisms on the nuclear ring clouds will be important in the future.

\section{Summary}
\label{sec:conclusions}

We presented high-resolution ($24\times18$~pc$^2$) \CO\ ALMA observations of the barred spiral galaxy NGC~613. We identified $356$ GMCs, $158$ of which are spatially and spectrally resolved, with a modified version of \textsc{cpropstoo}. We investigated the cloud properties and scaling relations and explored potential explanations for our findings. Our results can be summarised as follows:

\begin{enumerate}
\item The GMCs of NGC~613 have sizes ($15$ -- $75$~pc) that are comparable to, but molecular gas masses ($3\times10^5$ -- $5\times10^7$~\Msun), velocity dispersions ($2$ -- $36$~\kms) and molecular gas mass surface densities ($100$ -- $6000$~M$_\odot$~pc$^{-2}$) that are larger than those of the Milky Way disc and Local Group galaxy GMCs (\autoref{fig:histogram}). 

\item The GMCs sizes are similar across the different regions of the galaxy (nucleus, arcs, nodes and dust lanes), but the GMCs in the dust lanes have molecular gas masses, velocity dispersions and gas mass surface densities smaller than those of the GMCs in the other regions. The GMCs in the arcs tend to have gas masses and gas mass surface densities larger than those of the GMCs in the other regions.

\item The GMCs of NGC~613 have a steep size -- linewidth relation ($\sigma_{\mathrm{obs,los}}\propto R_{\mathrm{c}}^{0.77\pm0.11}$; \autoref{fig:larson_only}). 

\item The kinetic energy injection rate from the gas inflows driven by the large-scale bar is smaller than the turbulent energy dissipation rate, and the turbulence dissipation timescale is shorter than the cloud travel time, suggesting that additional sources of turbulence are required beside the gas inflows.

\item AGN feedback has a limited impact on the cloud turbulence, while stellar feedback, gas accretion and cloud-cloud collisions are plausible turbulence sources. Gas inflows into the nuclear ring from the dust lanes remain one of the likely explanations for the clouds' high velocity dispersions. However, quantitative analyses to assess the relative importance of the different mechanisms are required.
\item The GMCs of NGC~613 are marginally gravitationally bound ($\langle\alpha_{\mathrm{obs,vir}}\rangle\approx1.7$ with $X_{\rm CO}$ of the MW), and the GMCs in the arcs tend to have smaller virial parameters than the GMCs in the other regions.
\item The number of clouds decreases azimuthally downstream from the nodes within the nuclear ring (\autoref{sec:discussion_gmc_ring}). By tracking cloud disruption through GMC number statistics, we estimate the characteristic cloud lifetime to be between $3$ and $8$~Myr. This tends to be larger than the estimated timescales of cloud-cloud collisions, shear and/or stellar feedback ($\approx3$~Myr), suggesting that any of those could also contribute to the destruction of the clouds within the nuclear ring.
\end{enumerate}

As seen in this work, GMCs in nuclear rings clearly show distinct size-linewidth relations compared to those in other local environments when inspected in a similar manner. However, GMC properties and in particular the slopes of scaling relations can change depending on how individual clouds are defined \citep[e.g.][]{pan2015_gmc,dobbs2019_gmc}. Therefore, in a follow-up study, we will compare different cloud-finding algorithms to further validate our results (Choi et al. in prep.). Lastly, it is worth mentioning that we are currently analysing high spatial resolution numerical simulations of a galactic nuclear ring to gain physical insights into the high velocity dispersions of the GMCs, while also collecting more GMC observations of comparable quality for similar objects.

\section*{Acknowledgements}

We thank the anonymous referee for helpful and constructive comments and Dr. Chang-Goo Kim for useful discussions and comments. WC and AC acknowledge support by the National Research Foundation of Korea (NRF), grant Nos.\ 2022R1A2C100298212, and 2022R1A6A1A03053472. This work was also supported by National R\&D Program through the National Research Foundation of Korea (NRF) funded by the Korea government (Ministry of Science and ICT) (RS-2022-00197685). LL was supported by a Hintze Fellowship, funded by the Hintze Family Charitable Foundation, and by a DAWN Fellowship, funded by the Danish National Research Foundation under grant No.\ 140. MB was supported by UK Science and Technology Facilities Council (STFC) consolidated grant ``Astrophysics at Oxford'' ST/K00106X/1 and ST/W000903/1. TAD and IR acknowledges support from the STFC grants ST/S00033X/1 and ST/W000830/1. JG gratefully acknowledges financial support from the Swiss National Science Foundation (grant No.\ CRSII5 193826). This paper makes use of the following ALMA data: ADS/JAO.ALMA\#2017.1.01671.S. ALMA is a partnership of ESO (representing its member states), NSF (USA) and NINS (Japan), together with NRC (Canada), NSC and ASIAA (Taiwan) and KASI (Republic of Korea), in cooperation with the Republic of Chile. The Joint ALMA Observatory is operated by ESO, AUI/NRAO and NAOJ. This research has made use of the NASA/IPAC Extragalactic Database (NED), which is operated by the Jet Propulsion Laboratory, California Institute of Technology, under contract with the National Aeronautics and Space Administration.

\section*{Data availability}

The data underlying this article are available in the ALMA archive (\url{https://almascience.eso.org/asax/}) under project code 2017.1.01671.S. All higher-level data products will be shared upon reasonable request.

%%%%%%%%%%%%%%%%%%%%%%%%%%%%%%%%%%%%%%%%%%%%%%%%%%
%%%%%%%%%%%%%%%%%%%% REFERENCES %%%%%%%%%%%%%%%%%%

\bibliographystyle{mnras}
\bibliography{mnras_references}

%%%%%%%%%%%%%%%%%%%%%%%%%%%%%%%%%%%%%%%%%%%%%%%%%%
%%%%%%%%%%%%%%%%% APPENDICES %%%%%%%%%%%%%%%%%%%%%

\appendix

%%%%%%%%%%%%%%%%%%%%%%%%%%%%%%%%%%%%%%%%%%%%%%%%%%

% Don't change these lines
\bsp	% typesetting comment
\label{lastpage}
\end{document}